\documentclass[10pt,oneside]{amsart}

\usepackage{tikz}
\usetikzlibrary{automata, positioning, arrows}
\usepackage{longtable}
      \usepackage{amssymb}
\usepackage{amsfonts,amstext}
\usepackage{graphicx}
\usepackage{url}
\usepackage{comment}

\usepackage{amsmath,amstext,amssymb,amscd}
\usepackage{verbatim} 
\usepackage{mathtools}
      \usepackage{tabularx,booktabs}

\usepackage{amssymb,latexsym}
\usepackage{amsmath}
\usepackage{amsthm}
\usepackage{amssymb}
\usepackage{multicol}
\usepackage{makecell}

\theoremstyle{remark}

\theoremstyle{definition}

\usepackage{amsthm}
 \usepackage{multicol}
\allowdisplaybreaks
\theoremstyle{definition}
\newtheorem{example}{Example}
\newtheorem{definition}{Definition}

\usepackage{multicol}

      \makeatletter
      \def\@setcopyright{}
      \def\serieslogo@{}
     
      \makeatother

\begin{document}

\author{David McCune}
\address{David McCune, Department of Physics and Mathematics, William Jewell College, 500 College Hill, Liberty, MO, 64068-1896}
\email{mccuned@william.jewell.edu} 

\author{Adam Graham-Squire}
\address{Adam Graham-Squire, Department of Mathematical Sciences, High Point University, 1 University Parkway, High Point, NC, 27268}
\email{agrahams@highpoint.edu}

\title[Monotonicity Anomalies in Scottish Local Government Elections]{Monotonicity Anomalies in Scottish Local Government Elections}

\begin{abstract}
Single Transferable Vote (STV) is a voting method used to elect multiple candidates in ranked-choice elections.  One weakness of STV is that it fails multiple fairness criteria related to monotonicity and no show paradoxes. We analyze 1,079 local government STV elections in Scotland to estimate the frequency of such monotonicity anomalies in real-world elections, and compare our results with prior empirical and theoretical research about the rates at which such anomalies occur. In 62 of the 1079 elections we found some kind of monotonicity anomaly.  We generally find that the rates of anomalies are similar to prior empirical research and much lower than what most theoretical research has found.  The STV anomalies we find are the first of their kind to be documented in real-world multiwinner elections.
\end{abstract}



\maketitle

\section{Introduction}

The single transferable vote (STV) election procedure has been used for multiwinner elections in many countries since the early to mid-20th century.  For example, members of the Australian Senate have been elected using STV since 1948, and members of the D\'{a}il \'{E}ireann, the lower legislative house of the Irish legislature, have been elected using STV since 1921. In the 21st century the method has experienced a surge in interest and usage. Many municipalities in the United States currently use the single-winner version of STV, often referred to as instant runoff voting (IRV), for local elections. Such elections include city council races in Minneapolis, MN, Oakland, CA, and San Francisco, CA, as well as primary races for city office in New York City. IRV was even used for the 2020 US Presidential election in the state of Maine. In Scotland, STV has been used for multiwinner local government elections in council areas since 2007, and IRV has been used for a handful of single-winner elections.

While STV has its advantages as a voting method, such as its ability to achieve proportional representation in multiwinner elections, the method also has its drawbacks. One of the method's most serious weaknesses is that it is non-monotonic, where a candidate might be worse off receiving more support from voters (an \emph{upward monotonicity anomaly}), or a candidate might be better off receiving less support from voters (a \emph{downward monotonicity anomaly}). That is, the following scenario is possible when using STV: a candidate $X$ wins a seat but there exists a set of ballots such that if $X$ were moved up the rankings on these ballots, $X$ would not win a seat. Similarly, it is possible that $X$ does not win a seat but there exists a set of ballots such that $X$ would win a seat if they were moved down the rankings on these ballots. Other types of non-monotonicity are also possible. For example, it is possible that $X$ does not win a seat in an election but if fewer seats were available  then $X$ would win a seat (a \emph{committee size monotonicity anomaly}). Also, it is possible that a losing candidate $X$ would have won a seat if some of $X$'s supporters had abstained from voting in the election (a \emph{no-show anomaly}).

The purpose of this article is to investigate how often such anomalies occur in real-world elections. To that end, we collected and analyzed the freely available vote data from 1,079 Scottish local government elections, 30 single-winner and 1,049 multiwinner.  All elections used STV (or IRV) to elect a set of winners. For each type of monotonicity anomaly mentioned above, we wrote Python code that searched the ballot data from each of the Scottish elections to try to determine how many of the elections demonstrated the anomaly. Our general finding is that monotonicity anomalies occur rarely in these elections, occurring on the order of 1-4\% 
 for each type. As far as we are aware this paper is the largest empirical study of monotonicity to date, as the prior (mathematically-oriented) social choice literature has not analyzed this large database of Scottish STV elections.

\section{Previous literature on the frequency of monotonicity anomalies}\label{lit_review}

Previous literature regarding the frequency with which STV can produce monotonicity anomalies  mostly addresses only the single-winner upward case, and very little of this literature is empirical. One empirical analysis \cite{GZ} considered IRV elections in San Francisco and Alameda County, California between 2008 and 2016, as well as the 2009 mayoral election in Burlington, Vermont.  The study found an upward monotonicity anomaly rate of 0.74\% (1/135) of all IRV elections, 2.71\% (1/37) of IRV elections that went to at least a second round, and 7.7\% (1/13) of competitive three-candidate IRV elections. The most comprehensive empirical analysis of  IRV elections in the United States that went to a second round \cite{GMb} found anomaly rates of 2.2\% (upward), 1.6\% (downward) and 0.5\% (no-show). Additional empirical work tends to focus on a single election of interest, which does not provide insight on anomaly rates \cite{GM}, \cite{MM1}, \cite{ON}.

Semi-empirical research (i.e., research that does not have access to complete ballot preference data) finds small percentages of elections demonstrating anomalies when considering all elections, with estimates of zero \cite{B}, 0.028\% \cite{A}, 1.4\% \cite{Mi}, and 1.5\% \cite{G}. For extremely close elections, \cite{Mi} found that 33\% of elections demonstrate a monotonicity failure, and this percentage increases as elections become more competitive.  Both \cite{A} and \cite{B} address multiwinner STV elections, but \cite{A} uses poll data in the absence of complete preference data and considers only very restricted kinds of monotonicity anomalies, and the methodology in \cite{B} is not clear. In a semi-empirical analysis, \cite{KST} found that 20\% of past French presidential elections likely demonstrated a monotonicity failure under the voting method of plurality runoff, which is similar to IRV.

Theoretical research into three-candidate IRV elections tends to find a higher frequency of upward anomalies, although the prevalence varies depending on the assumptions of the chosen model of voter behavior and the closeness of the election.  Estimates that 1.76\% to 4.51\% of all elections would demonstrate upward anomalies are found in \cite{LDB}, where the percentage depends on which model of voter behavior is used. Between 4.5\% and 6.9\% was found in \cite{Q}, whereas \cite{PT} finds a frequency of less than 1\%.  Using a different model of voter behavior and a broader definition of monotonicity, \cite{Q} found that the percentage of elections demonstrating anomalies tends to 100\% as the number of candidates increases. In elections where the top three candidates all receive more than 25\% of the first-place vote, estimates range from as low as 10\% \cite{Mi} to 51\% in highly competitive elections where the top three candidates are in a virtual tie \cite{ON}.  

Some theoretical research has also examined the prevalence of downward and no-show anomalies in three-candidate IRV elections.  For downward anomalies, estimates for a lower bound range from 1.97\% \cite{LM} to 3.8\% \cite{Mi}. For no-show anomalies, \cite{PT} found rates of 0.38\% to 0.47\%, and \cite{LM} found rates about 10 times higher, between 4.1\% and 5.6\%.  The former used a spatial model, and the latter utilized the impartial anonymous culture and impartial culture models.  In empirical research, \cite{GZ} found a rate of 0\% for no-show anomalies in the 135 IRV elections analyzed. There has been no prior theoretical analysis of the frequency of committee size anomalies.

As far as we are aware, there have been no prior documented monotonicity anomalies of any kind in real-world multiwinner elections, where by ``documented'' we mean that full preference data is available and a set of ballots can be found which demonstrate the given anomaly. The reason for the lack of examples is that the database of Scottish elections is the first large set of multiwinner elections with available preference data which has been searched for monotonicity anomalies. All prior documented instances of monotonicity anomalies have occurred in single-winner IRV political elections in the United States, which are listed below.

\begin{itemize}
\item The 2009 mayoral election in Burlington, VT, which demonstrated an upward anomaly \cite{Mi}, \cite{ON}.
\item The 2020 board of supervisors election in the seventh ward of San Francisco, CA, which demonstrated a downward anomaly \cite{GMb}.
\item The 2021 city council election in the second ward of Minneapolis, MN, which demonstrated upward and downward anomalies \cite{MM1}.
\item The August 2022 Special Election for the US House of Representatives in Alaska, which demonstrated upward and no-show anomalies \cite{GM}.
\item The 2022 school director election in district 4 of Oakland, CA, which demonstrated upward and downward anomalies \cite{Mc}.
\end{itemize}

Our results (Table \ref{election_summary}) significantly increase the number of documented monotonicity anomalies in real-world elections.

\section{Preliminaries: Single Transferable Vote and Monotonicity Anomalies}\label{preliminaries}

The Scottish elections we study use the method of STV to choose the set of election winners. There are different voting methods which can be classified as \emph{STV}; we use the term ``STV'' to refer only to the Scottish STV rules, which we outline below. 

Let $n$ denote the number of candidates in an election and let $S$ denote the size of the winner set, which equals the number of available legislative seats. In an STV election, each voter casts a preference ballot where the voter provides a preference ranking of the candidates. In Scottish elections voters are not required to provide a complete ranking and thus it is common for voters to rank only a subset of the candidates, leaving some candidates off their ballots. The ballots are combined into a \emph{preference profile}, which provides a count of how many different kinds of ballot were cast; the preference profile of each election is the data we collected and analyzed. Table \ref{example_profile} shows an example of a preference profile in an election with 501 voters and $n=4$ candidates $A$, $B$, $C$, and $D$. The table shows that 19 voters rank $A$ first, $B$ second, and leave $C$ and $D$ off the ballot; the other numbers across the top row convey similar information about the number of voters who cast the corresponding ballot. When discussing a given ballot we use the notation $\succ$ to denote that a candidate is ranked immediately above another candidate, so that 41 people cast the ballot $A \succ B \succ C \succ D$, for example. An \emph{election} is an ordered pair $(P,  S)$ where $P$ is a preference profile. STV takes an election as input and outputs a winner set, which we denote $W(P, S)$.

\begin{table}
\begin{tabular}{l|c|c|c|c|c|c|c|c|c|c|c|c|c}
Num. Voters&19&41&60&15&73&51&19&57&12&40&8&47&59\\
\hline
1st Choice& A& A & A  & A & B & B & B & C & C & C & D & D & D\\
2nd Choice &B& B & C & D & C & A & D & A & B & D & A & C & B\\
 3rd Choice & & C & D &    & A & D & C &  & A & B & C & B & \\
  4th Choice & & D &    &     &    & C & A &  & D & A &  & &\\

\end{tabular}

\caption{An example of a preference profile with 501 voters.}
\label{example_profile}
\end{table}

It is difficult to provide a complete definition of STV in a concise fashion. Therefore, we provide a high level description which we illustrate using examples with the preference profile in Table \ref{example_profile}. The formal description of the rules can be found at \url{https://www.legislation.gov.uk/sdsi/2007/0110714245}.

The method of STV proceeds in rounds. In each round, either a candidate earns enough votes to be elected or no candidate is elected and the candidate with the fewest (first-place) votes is eliminated. The number of votes required to be elected is called the \emph{quota}, and is calculated by \[\text{quota } = \left\lfloor \frac{\text{Number of Voters}}{S+1}\right\rfloor +1.\]

If no candidate reaches quota in a given round then the candidate with the fewest first-place votes is eliminated, and this candidate's votes are transferred to the next candidate on their ballots who has not been elected or eliminated. If a candidate reaches quota, that candidate is elected and the votes they receive above quota (\emph{surplus votes}) are transferred in a fashion similar to that of an eliminated candidate, except the surplus votes are transferred in proportion to the number of ballots on which each other candidate appears.  To explain how these transfers work, suppose candidate $A$ is elected with a total of $a$ votes and a surplus of $A_s$ votes (so that $A_s = a - $ quota), and candidate $B$ is the next eligible candidate on $b$ of these ballots. Rather than receive $b$ votes from the election of $A$ candidate $B$ receives $(A_s/a)b$ votes, resulting in a fractional vote transfer. The method continues in this fashion until $S$ candidates are elected, or until some number $S'<S$ of candidates have been elected by surpassing quota and there are only $S-S'$ candidates remaining who have not been elected or eliminated.

We illustrate this description using the preference profile in Table \ref{example_profile} and seat values of $S=1$ and $S=2$.

\begin{example}\label{first_example}
When $S=1$ the quota is $\lfloor 501/2 \rfloor +1 = 251$ and a candidate must receive a majority of votes to win. No candidate initially receives a majority of first-place votes and thus $C$, the candidate with the fewest first-place votes, is eliminated. As a result 57 votes are transferred to $A$, 12 to $B$, and 40 to $C$, as displayed in the vote totals for the next round of votes in the left side of Table \ref{first_STV_example}. We refer to such a table as a \emph{votes-by-round} table, and often display it in lieu of a preference profile when the profile is too large to display. None of the remaining candidates have reached quota and thus $D$, who now has 154 votes, is eliminated, causing 56 votes to transfer to $A$ and 146 votes to transfer to $B$. The STV method declares $B$ the winner, as they have now surpassed quota. Thus, $W(P,1)=\{B\}$. 

A transfer of surplus votes never occurs when $S=1$. This changes when $S=2$, as shown in the right table of Table  \ref{first_STV_example}. In this case the vote totals in the first two rounds are identical to the $S=1$ case because no candidate achieves quota in the first round; however, $A$ surpasses quota in the second round and their 24 surplus votes must be transferred. Since $C$ has been eliminated, $60(24/192)=7.5$ votes are transferred to $B$, $75(24/192)=9.375$ votes are transferred to $D$, and $57(24/192)=7.125$ votes are removed from the election because the 57 ballots of the form $C \succ A$ do not indicate which candidate should these receive votes should $A$ be elected or eliminated. Therefore, in the third round $B$ has 162.500 votes and $D$ has 163.375. $B$ is eliminated, causing $D$ to surpass quota with 233.375 votes. Thus, $W(P,2)=\{A,D\}$.

Note that if $D$ were not to appear on any of the ballots that are transferred when $B$ is eliminated then $D$ would finish with only 163.375 votes,  4.625 votes shy of quota. Since there is still one seat left to fill, $D$ would be elected because they are the only candidate left, and this would be an example where a candidate wins without achieving quota.

\begin{table}
\begin{tabular}{cccc}
\begin{tabular}{c|c|c|c}

\multicolumn{4}{c}{$S=1$, quota $=$ 251}\\
\hline
\hline
Cand.& \multicolumn{3}{c}{Votes By Round}\\
\hline
$A$& 135&192&200\\
$B$& 143& 155&\textbf{301}\\
 $C$& 109& &  \\
$D$& 114&154&\\

\hline
\end{tabular}

&&&
\begin{tabular}{c|c|c|c|c}

\multicolumn{5}{c}{$S=2$, quota $=$ 168}\\
\hline
\hline
Cand.& \multicolumn{4}{c}{Votes By Round}\\
\hline
$A$& 135&\textbf{192}&\\
$B$& 143& 155&162.500&\\
 $C$& 109& &  &\\
$D$& 114&154&163.375& \textbf{233.375}\\

\hline
\end{tabular}

\end{tabular}
\caption{The left (respectively right) table shows the vote totals for each candidate by round, and eventual STV winners, for $S=1$ (respectively $S=2$) seats. A bold number represents when a candidate is elected.}
\label{first_STV_example}

\end{table}

\end{example}

As mentioned in the introduction, we are interested in four types of monotonicity anomaly that can occur in STV elections. We now define each type, focusing on the multiwinner context since 97\% of the elections in our database satisfy $S>1$.  Because we are concerned with how these anomalies manifest in our database of actual elections and because our work with these elections never produces ties, our definitions assume a unique winner set. A careful theoretical treatment of these anomalies, such as what appears in \cite{EFSS}, must take ties into account and thus articles like \cite{EFSS} treat STV as a set-valued method that can output multiple sets of winners, and defines the various monotonicity anomalies accordingly. We avoid the issue of ties, and the corresponding technical notation, due to the empirical nature of our work.

Our first type of monotonicity, which we term \emph{committee size monotonicity} following  terminology in \cite{EFSS}, was first introduced in \cite{S}. Committee size monotonicity requires that when we increase the number of seats available, every candidate who won a seat under the smaller seat size still wins a seat under the larger seat size. 

\begin{definition}
\textbf{(Committee Size Monotonicity)} Given an election $(P, S)$, for any $1 \le i < S$ we have $W(P,i) \subseteq W(P,S)$. 
\end{definition}

An election $(P,S)$ for which there exists $1\le S' <S$ such that $W(P,S') \not\subset W(P,S)$ is said to demonstrate a \emph{committee size monotonicity anomaly}.  Such an anomaly is found in Example \ref{first_example}: note that $W(P, 1)=\{B\}$, which is not a subset of $W(P, 2)= \{A, D\}$. It seems paradoxical $B$ is simultaneously the ``best'' single candidate when $S=1$, but not in the ``top half'' of candidates when $S=2$.

One of the reasons monotonicity anomalies are of interest to social choice theorists is that anomalies can demonstrate ``harm'' toward a political candidate or some voters, and that harm seems paradoxical. In this example, it is understandable if candidate $B$, and voters who prefer that $B$ receive a seat, feel treated unfairly by the outcome. In addition to candidates and voters feeling harmed, in partisan elections (i.e., elections in which candidates belong to a political party) it is also possible for political parties to be harmed. Suppose in this example $B$ belongs to the Scottish Labour Party but $A$ and $D$ belong to the Scottish Conservative Party. Then Labour loses their only seat in moving from $S=1$ to $S=2$, and thus the party is harmed as well. Most of the previous literature on monotonicity anomalies implicitly studies non-partisan elections, choosing to focus only on the candidates, and sometimes the voters, affected by an anomaly. Since our study concerns partisan Scottish elections, we also discuss harm to political parties when presenting our results.

We note that an empirical analysis of committee size paradoxes has limitations, in that we cannot know if voters would vote substantially differently if the number of seats available were different. If Example \ref{first_example} were a real-world election with $S=2$, we would need to conduct high quality polls to know if $B$ would be the IRV winner when $S=1$. We do not have access to such poll data for the Scottish elections and thus we use the definition of committee size monotonicity from the previous literature, which assumes the same underlying vote data for each choice of $S$.

We now define the other three types of monotonicity, which have been  studied primarily in a single-winner context in which it is assumed that each voter casts a ballot with a complete ranking of the candidates. Adapting these definitions to a real-world multiwinner context in which voters often cast partial ballots is not straightforward. First, we state how we handle partial ballots. We adopt the \emph{weak order model} \cite{PPR} wherein we assume that a voter who casts a partial ballot is indifferent among candidates left off the ballot, all of which are ranked beneath candidates that appear on the ballot. We use only the preference information provided by the voter, and choose not to try to complete partial ballots using statistical inference. In this way we are similar to an office of elections, which does not infer any information on a ballot beyond what a voter communicated. As discussed in \cite{PPR} there are other ways to process partial ballots, but empirical studies regarding STV tend to interpret partial ballots as we do (see \cite{GZ}, \cite{KGF}, \cite{MM2}), although similar studies which also use real-world elections to generate simulated elections often handle partial ballots in a variety of ways (see \cite{PPR}, for example).

Informally, upward monotonicity states that a candidate who wins a seat should not become a loser by gaining voter support, where that extra support consists of shifting the winning candidate up the rankings on some ballots and leaving the relative rankings of the other candidates unchanged. Because we use the weak order model for partial ballots, ``shifting a winner up the rankings'' includes scenarios where the winning candidate does not appear on the actual ballots and we place that winner at the first ranking on these ballots, shifting all other candidates down one ranking. We note that we choose the term ``upward monotonicity'' to accord with the literature for the single-winner case; this notion of monotonicity is also referred to as \emph{candidate monotonicity} in \cite{EFSS}.

\begin{definition}
\textbf{(Upward Monotonicity)} Given an election $(P, S)$, let $X \in W(P,S)$ and let $\mathcal{B}$ be a set of ballots from $P$. If we construct a new preference profile $P'$ from $P$ by moving $X$ to a higher position in the ballots from $\mathcal{B}$ but leave unchanged the relative positions of all other candidates on the ballots from $\mathcal{B}$ then $X \in W(P', S)$.

\end{definition}

An election is said to demonstrate an \emph{upward monotonicity anomaly} if there exists a winning candidate $X$ and a set of ballots $\mathcal{B}$ such that moving $X$ to a higher position on the ballots from $\mathcal{B}$, but leaving the relative positions of the other candidates unchanged, creates a preference profile in which $X$ loses.

Informally, downward monotonicity states that a candidate who does not win a seat  should not become a winner by losing voter support, where that lost support consists of shifting the candidate down the rankings on some ballots and leaving the relative rankings of the other candidates unchanged. Because of partial ballots, downward monotonicity is more difficult to define in a real-world context. For example, suppose candidate $A$ does not win a seat but $A$ would win a seat if we take 10 ballots with $A$ ranked first and no other candidates listed on the ballot (we refer to such ballots as \emph{bullet votes} for $A$) and change those ballots to bullet votes for $B$. Under the weak order model, shifting $B$ up the rankings in such a manner changes the relative ordering of the candidates besides $A$, and thus such an outcome would not count as a violation of downward monotonicity under a traditional definition. However, this scenario  fits the spirit of a downward monotonicity violation. To deal with this issue of partial ballots, we adapt the classical single-winner definition of downward monotonicity into strong and weak forms, where  the strong form insists that the relative rankings of candidates besides the affected losing candidate are unchanged (similar to the classical notion of downward monotonicity), whereas the weak form allows for situations in which we change bullet votes.

\begin{definition}
\textbf{(Downward Monotonicity)} Given an election $(P, S)$, let $X \not\in W(P, S)$ and let $\mathcal{B}$ be a set of ballots from $P$ such that $X$ appears on all ballots in $\mathcal{B}$. 
\begin{itemize}
\item \textbf{Strong Downward Monotonicity}: If we construct a new preference profile $P'$ from $P$ by moving $X$ to a lower position in the ballots from $\mathcal{B}$ but leave unchanged the relative positions of all other candidates on the ballots from $\mathcal{B}$ then $X \not\in W(P', S)$.
\item \textbf{Weak Downward Monotonicity}: Let $\mathcal{B}_1$ and $\mathcal{B}_2$ be a partition of $\mathcal{B}$ such that $\mathcal{B}_2$ consists of bullet votes for $X$. If we construct a new preference profile $P'$ from $P$ by moving $X$ to a lower position in the ballots from $\mathcal{B}_1$ but leave the relative positions of all other candidates on the ballots from $\mathcal{B}_1$ unchanged, and we change all ballots in  $\mathcal{B}_2$  to bullet votes for $Y$ or to ballots of the form $Y \succ X$ for some candidate $Y\neq X$, then $X \not\in W(P', S)$.

\end{itemize}

\end{definition}

A \emph{downward monotonicity anomaly}, either strong or weak, is defined similarly to an upward monotonicity anomaly.

 When $S=2$, the election with the preference profile in Table \ref{example_profile} contains both an upward and a strong downward monotonicity anomaly. To demonstrate the upward anomaly, observe that if six voters who cast the ballot $D\succ A \succ C$ move $A$, who is a winner in the original election, up one ranking so that the 6 ballots become $A \succ D \succ C$, then $A$ no longer wins a seat. As illustrated in the left votes-by-round table of Table \ref{up_down_anomaly}, even though $A$ receives more votes initially, shifting $A$ up on those 6 ballots causes $D$ to be eliminated first instead of $C$ and the winner set changes from $\{A,D\}$ to $\{B,C\}$. That is, as a result of 6 voters being persuaded that $A$ is their favorite candidate rather than their second-favorite, $A$ becomes a losing candidate because the order of elimination/election changes. Note that for this outcome to count as an anomaly we simply need $A$ to drop from the winner set; the simultaneous removal of $D$ is an unfortunate side effect for this candidate, but if moving $A$ up on some ballots causes $D$ to lose but $A$ remains a winner, we do not say that an anomaly occurred.

 \begin{table}
\begin{tabular}{cccc}
\begin{tabular}{c|c|c|c}

\multicolumn{4}{c}{$S=2$, quota $=$ 168}\\
\hline
\hline
Cand.& \multicolumn{3}{c}{Votes By Round}\\
\hline
$A$& 141&143&151.58\\
$B$& 143& \textbf{202}&\\
 $C$& 109& 156 & \textbf{171.49} \\
$D$& 108&&\\

\hline
\end{tabular}

&&&
\begin{tabular}{c|c|c|c}

\multicolumn{4}{c}{$S=2$, quota $=$ 168}\\
\hline
\hline
Cand.& \multicolumn{3}{c}{Votes By Round}\\
\hline
$A$& 135&143&150.29\\
$B$& 137& \textbf{196}&\\
 $C$& 115& 162& \textbf{174.29} \\
$D$& 115&&\\

\hline
\end{tabular}
\end{tabular}
\caption{The left (respectively right) table demonstrates an upward (respectively downward) monotonicity anomaly for the election $(P,2)$ from Example \ref{first_example}.}
\label{up_down_anomaly}

\end{table}

To demonstrate a strong downward monotonicity anomaly, suppose 6 voters who cast the ballot $B\succ C \succ A$ in the original election cast the ballot $C\succ B \succ A$ instead, moving $B$ down one ranking. As illustrated in the right example of Table \ref{up_down_anomaly}, $D$ is eliminated first and the winner set is $\{B,C\}$ for the modified election. If $B$ were moved down one ranking on this handful of ballots, $B$ would have been an election winner rather than a loser.

We now define our final type of monotonicity, participation monotonicity, and its corresponding type of anomaly, a no-show anomaly (this is also sometimes referred to as an \emph{abstention paradox}). Informally, participation monotonicity requires that voters are better off casting ballots than abstaining from the election. This is succinctly stated in \cite{KMNZ}: ``it should always be better to vote honestly than not to vote at all.'' The notion of a no-show anomaly has been formally defined in different ways in the context of single-winner elections.  For example, \cite{FN} states (harkening back to the original definition in \cite{M}), ``The no-show paradox occurs whenever a group of identically minded voters is better off abstaining than by voting according to its preferences.'' In such a definition, the group of voters affected by the paradox must all cast the exact same ballot.  Other definitions relax this assumption. Consider the definition from \cite{HP}: ``if a candidate $x$ is the winner in an initial election, then if we add to that scenario some new voters who rank $x$
above $y$, then the addition of these new voters should not make $y$ the winner.'' Under this definition, the voters affected by the anomaly need not cast identical ballots, they merely must agree that they prefer $x$ to $y$. 

We are unaware of previous attempts to define participation monotonicity in a multiwinner context in which voters cast preference ballots. Definitions have been proposed for multiwinner elections which do not use preference ballots (see \cite{SFF}, for example), but such definitions do not easily translate to the STV setting. We choose to adapt the definition from \cite{HP}, but multiwinner elections contain subtleties which complicate attempts to formalize the sentiment ``it should always be better to vote honestly than not to vote at all.'' The reason is that, as argued in \cite{R}, a voter's preferences about winner sets cannot always be distilled into a preference ranking of the individual candidates.  For example, suppose in a three-seat election a voter casts the ballot $A\succ B \succ C \succ D \succ E\succ F$. From this ranking it is clear that the voter prefers a winner set of $\{A,B,C\}$ to $\{D,E,F\}$, but does this voter prefer $\{A,C,F\}$ to $\{B,C,E\}$? Given only the voter's preference ranking of the candidates, we cannot say. A more pertinent question when trying to define a no-show anomaly is: does this voter prefer $\{A,B,D\}$ to $\{A,B,E\}$? Suppose that when the voter participates in the election the winner set is $\{A,B,E\}$ but when they abstain the winner set is $\{A,B,D\}$; is the voter necessarily worse off when they cast a ballot? We choose to say the answer is Yes; however, it is conceivable that the voter would prefer $\{A,B,E\}$ to $\{A,B,D\}$, perhaps because of the group dynamics of the three candidates.

In addition to the concerns outlined above, there are computational challenges when searching for no-show anomalies in actual data. For these reasons, we prefer to focus on winner changes among only the two candidates $x$ and $y$ from the definition in \cite{HP}. Thus, our definition of a no-show anomaly insists that if voters who prefer $x$ to $y$ abstain rather than vote, the only change to the winner set is that $x$ replaces $y$. Other definitions, either more or less restrictive, are also sensible.

\begin{definition}
\textbf{(Participation Monotonicity)} Let $(P, S)$ be an election, with $X \not \in W(P, S)$ and $Y \in W(P, S)$. Let $\mathcal{B}$ be a set of ballots on which $X$ is ranked higher than $Y$. Then if we remove the ballots in $\mathcal{B}$ from the election, it should not be the case that the resulting winner set is $(W(P, S) -\{Y\}) \cup \{X\}$.
\end{definition}

A \emph{no-show anomaly} is said to occur in an election $(P, S)$ if there exists $X \not\in W(P, S)$, $Y \in W(P, S)$, and a set of ballots $\mathcal{B}$ on which $X$ is ranked higher than $Y$ such that if the ballots from $\mathcal{B}$ were removed from the preference profile then $X$ replaces $Y$ in the winner set.

Given the potential ambiguity about whether a set of voters truly is better off not voting, when searching for no-show anomalies we look for instances of the anomaly that are unambiguous. Specifically, we only report instances in which candidate $Y$ is not ranked in the top $S$ candidates. Furthermore, we try to find instances in which candidate $X$ is ranked in the top $S$ candidates and, if possible, we remove a set of identical ballots to create the anomaly. Such constraints seem like a good way to generalize the various notions of a no-show paradox from the single-winner literature. The Appendix provides the details of the no-show anomalies we report so that the reader can judge how unambiguous the results are.

Our running example $(P,2)$ demonstrates a no-show anomaly: if 35 voters who cast the ballot $B \succ C \succ A$ are removed from the election, creating the preference profile $P'$, then $W(P',2)=\{A,C\}$. These 35 voters prefer $C$ to $D$, yet when they cast a ballot $D$ is a winner, and when they abstain $D$ is replaced by $C$ in the winner set. In this example the voters removed from the election cast identical ballots but for our definition of a no-show anomaly, it is only relevant that the voters prefer $C$ to $D$. Furthermore, this seems like an unambiguous instance of a no-show anomaly, as these voters rank $C$ in their top two and thus presumably they truly are worse off when $D$ (who does not appear on their ballots) replaces $C$ in the winner set.

To conclude this section we note that these four types of monotonicity are logically independent, in the sense that an election which contains an upward anomaly may not contain a downward or a committee size anomaly, for example. An election such as our running example which demonstrates all four types of anomaly is most likely extremely rare. We found no examples of a Scottish election that exhibits all four anomalies, although four 
 elections demonstrate three of the four. Before providing our results about the frequency of monotonicity anomalies in real-world elections, we discuss our sources of data and how we searched the data for anomalies.

\section{Data Sources: Scottish Local Government Elections}\label{data_source}

For the purposes of local government, Scotland is partitioned into 32 council areas, each of which is governed by a council. The councils provide a range of public services that are typically associated with local governments, such as waste management, education, and building and maintaining roads. The council area is divided into wards, each of which elects a set number of councilors to represent the ward on the council. The number of councilors representing each ward is determined primarily by the ward's population, although other factors play a role\footnote{For complete details about how the number of councilors for a ward is determined, see \url{https://boundaries.scot/reviews/fifth-statutory-reviews-electoral-arrangements}.}. Every five years each ward holds an election in which all seats available in the ward are filled using the method of STV.

Every Scottish ward has used STV for local government elections since 2007. Preference profiles from the 2007 elections are difficult to obtain; we contacted several council election offices and either received no response or were told that the 2007 data is not available. Thus there are no elections from 2007 in our database. We obtained preference profile data for the 2012 and 2017 ward elections from the Local Elections Archive Project \cite{T}, although some of this data is still available on various council websites. We obtained data for the 2022 preference profiles from the  council websites.

In addition to the regularly scheduled local government elections which occur on a five-year cycle, council areas sometimes hold off-schedule by-elections to fill a seat that is open due to the death or resignation of a councilor. These by-elections are almost always single-winner IRV elections. The data for many of these elections is not available because some councils  hand-count these ballots, not using the STV tabulation software that is used for the regularly-scheduled elections. We obtained preference profiles for the available by-elections from various council websites, and by request from several council election offices.

In all, we collected the preference profile data of 1,079 STV elections, 30 single-winner and 1,049 multiwinner. While we would prefer to have preference data from all Scottish local government elections, including 2007 elections and all off-schedule by-elections, the database we use is large enough to make robust conclusions about the frequency of monotonicity anomalies in real-world STV elections.

As mentioned in Section \ref{lit_review}, this collection of actual ballot data is what sets our study apart from most of the prior empirical and semi-empirical research on monotonicity anomalies. For each election in our database we have a complete record of the preference ranking of candidates expressed by each  voter, which means that we do not need to rely on surveys or other such tools to search for monotonicity anomalies. When we detect an anomaly, we can provide an exact set of ballots, and (in the case of an upward or downward anomaly) how to alter the ballots, to demonstrate an anomaly.
\begin{table}

\begin{tabular}{l|c c c c c}
Num. Seats & 1&2&3&4&5\\
\hline
Num. Elections & 30 & 5 & 549 & 492 & 3\\
\end{tabular}

\caption{The number of elections in the database of 1,079 elections with the given number of seats.}
\label{num_seats}
\end{table}

\begin{table}

\begin{tabular}{l|cccccccccccc}
Num. Cands     & 3&4&5&6&7&8&9&10&11&12&13&14\\
\hline
Num. Elections& 3&39&119&212&289&205&113&63&22&8&5&1\\
\end{tabular}
\caption{The number of elections in the database of 1,079 elections with the given number of candidates.}
\label{num_cands}
\end{table}

We conclude this section  by providing basic information about the number of voters, candidates, seats and voter behavior in these Scottish elections. Across all elections the minimum number of voters\footnote{When we refer to a ``number of voters,'' we mean the number of voters who cast a valid ballot. Ballots with errors are not counted in these elections.} in an election is 661, the maximum is 14,207, and the median is 4,790. Thus, the electorates under consideration are not tiny, but the size of an electorate in these Scottish elections tends to be much smaller than electorates in many other publicly accessible databases of elections that use preference ballots. For example, the city of Minneapolis, Minnesota uses IRV to elect a single city councilor from each of its 13 wards. In the 2021 Minneapolis city council elections\footnote{The vote data for these elections can be found at \url{https://vote.minneapolismn.gov/results-data/election-results/2021/}.} the median number of voters across the wards was 11,326, more than double the median from the Scottish elections. Electorates from other American IRV elections in places such as New York City or the state of Maine tend to be much larger.  

Table \ref{num_seats} (resp. \ref{num_cands}) shows a breakdown of the number of elections by number of seats (resp. candidates). The number of seats for elections in the database tends to be 3 or 4; there was no election with $S>5$. The number of candidates ranges from 3 to 14, although the majority of elections have 6, 7, or 8 candidates. 

In Scottish local government elections voters are not required to provide a complete ranking of all the candidates, and  thus many of the ballots contain only a partial ranking (often referred to as \emph{ballot truncation}).  When we process the ballot data we assume that a voter prefers any candidate ranked on their ballot to any candidate not ranked on their ballot and we make no inference as to how the voter would have ranked candidates left off their ballot. It is possible that our results would change if the ballots were processed differently; we handle the ballots as we do because we prefer to consider precisely the ranking information provided by the voters.  We note that ballot truncation is more the norm than an aberration in Scottish elections.  Specifically, the average voter casts a ballot which ranks fewer candidates than seats to be elected, and much less than the number of available candidates.  Table \ref{avg_length} shows the average length of a ballot for elections with a given number of seats.  We note as well that the median length of a ballot was 3 in almost all situations.  One exception is that the median length in 2-seat elections was only 2, and that was also the only situation in which voters on average ranked more candidates than the available number of seats.  The 2-seat elections only comprise 5 of the 1049 
 multiwinner elections, however.  To get a sense of the relationship between average ballot length and the number of candidates, Table \ref{avg_length_numCand} shows that as the number of candidates increases in a 4-seat election, the average ballot length also generally increases.  However, the growth is quite slow--in elections with 7 or more candidates, the average voter ranks less than half of the candidates.  The median ballot length in 4-seat elections was 3 for all numbers of candidates, except in 13- and 14-seat elections where the medians were 3.25 and 3.5 (respectively).

\begin{table}
\begin{tabular}{l|cccc}
Number of Seats     & 2& 3&4&5\\
\hline
Avg. Ballot Length& 2.59&2.98&3.30&3.77\\
\end{tabular}
\caption{Average number of rankings for the given number of seats in an election.}
\label{avg_length}
\end{table}

\begin{table}
\begin{tabular}{l|cccccccccc}
Number of Candidates     &5&6&7&8&9&10&11&12&13&14\\
\hline
Avg. Ballot Length&2.84&3.02&3.16&3.32&3.42&3.73&3.79&4.03&3.97&4\\
\end{tabular}
\caption{Average number of rankings with the given number of candidates in 4-seat elections.}
\label{avg_length_numCand}
\end{table}

\section{Methodology: How We Search for Monotonicity Anomalies}\label{methodology}

In this section we provide a high level description of the code we created to search for monotonicity anomalies. The code is available at \cite{GM_Git}, and is adapted from programs used in \cite{GZ}.

Searching for committee size anomalies is straightforward: calculate $W(P,S')$ for $1\le S' <S$ and check if $W(P,S') \subset W(P,S)$. If an election contains a committee size anomaly then such code definitely finds it.

Searching for the other types of monotonicity anomaly in an election is much more difficult, as the code must search for a set of ballots which demonstrate the given anomaly. Unless $S=1$ and $n=3$ (which occurs in none of our elections) there are no known necessary and sufficient conditions for an election to demonstrate a given anomaly, and therefore if an anomaly exists we cannot guarantee that our code will find it. We use two different kinds of program to search for anomalies, as outlined below. The first type of code is sophisticated in that it analyzes the votes-by-round table to try to engineer an anomaly by changing the order in which candidates are elected or eliminated. The second type of code uses brute force by systematically changing batches of ballots.

The first type of program works as follows. At each round of the election, the programs look for modifications to the preference profile (raising or lowering a candidate’s ranking, or eliminating certain ballots) that could change the order of elimination or candidates being elected in the original election, and then the programs check to see if the modified profile would result in appropriately different winners.  We provide a more detailed description of the upward monotonicity program; the downward and no-show programs are conceptually similar. The upward monotonicity program first runs the original STV election and calculates the winner set $W(P,S)$ and the set $E$ of eliminated candidates, in order of elimination.  Let $\mathcal{C}$ denote the set of candidates in the election and $E_1$ be first eliminated candidate, $E_2$ the second eliminated, etc.  

The program then has two phases, one where it looks for anomalies with changes in the elimination order, and the other where it looks for anomalies related to changes in the manner in which candidates are elected. To look for changes in the elimination order, the code chooses a winner $W_m \in W(P,S)$, and a candidate $C_i$ in $\mathcal{C}-\{W_m, E_1\}$.  The program checks for ballots with $C_i$ listed first where the following would happen: $W_m$ could be raised higher in enough ballots so that $C_i$ would be eliminated before $E_1$, without first making $W_m$ surpass quota.  If such ballots exist, the program shifts $W_m$ to the top of all such ballots and reruns the election with the modified profile $P’$.  If $W_m$ is not in $W(P’, S)$, then the program reports an anomaly.  The program then reverts back to the original profile $P$ and checks all other $C_k$ for a given $W_m$, then chooses a different $W_j$ and repeats the process until all $W_m$ and $C_i$ have been exhausted at the level of $n$ candidates. At this point, the program eliminates candidate $E_1$ to get a new profile $P_{n-1}$, and repeats the process above for the second eliminated candidate $E_2$, remaining winners $W_m$, and remaining candidates $C_i$.  The program continues eliminating candidates and checking all possible changes of elimination order until all eliminated candidates are exhausted. If an anomaly is reported at this stage then it is possible that the program has returned a false positive, which occurred a few times.  

In the second phase to look for changes in the seat order, the program reverts back to the original election, and then runs the election to the point where a candidate $W_1$ is elected.  The code then tries to move enough ballots so that $W_1$ is $not$ elected in that round of the election by doing the following: it chooses a different winner of the election, $W_2$, who was elected after $W_1$, and finds ballots where $W_2$ is ranked second and $W_1$ is ranked first, and modifies the ballots so that $W_2$ moves above $W_1$, thus reducing the number of first-place votes that $W_1$ has.  Once those ballots are modified, the code reruns the election to see if $W_2$ is still a winner.  If not, the code reports an upward monotonicity anomaly.  If so, the code reverts back to the original election data and repeats the process with all other such combinations of winners.

The second type of program uses brute force to make small, incremental changes in the ballot data. To search for upward anomalies, the code works as follows. For each winner $W_i$ and each other candidate $B$, the code finds ballots on which $B$ is ranked first and, ballot-by-ballot, shifts $W_i$ up to the first ranking. The code starts with ballots on which $W_i$ ranked second, then moves on to ballots on which $W_i$ is ranked third, etc. The code also tries other orderings of ballots to use. For example, we also ran code which started the ballot swaps with ballots on which $B$ is ranked first and $W_i$ is ranked second, then moved on to bullet votes for $B$. If the code finds a set of such ballot changes which produces an election in which $W_i$ is no longer a winner, an anomaly is reported. The downward and no-show programs are implemented similarly.

The brute-force code found most of the upward and downward anomalies found by the more sophisticated code, and the sophisticated code found all anomalies reported by the brute-force code. Thus, for the case of downward and upward anomalies the brute-force code turned out to be redundant. The case of no-show anomalies is much different: the brute-force code found several no-show anomalies not found by the first kind of code. The reason is that, as outlined below, no-show anomalies can occur in STV elections without changing the order of election or elimination of candidates (this is in stark contrast to the single-winner case, in which any type of anomaly occurs by changing the order of elimination).

While we cannot guarantee that we have found all anomalous elections, we did the following to test and double-check our work:
\begin{itemize}
\item All programs were tested on elections we created that had different anomalies to make sure the programs would find different varieties of how the anomalies present.
\item All anomalies reported in this paper were discovered by our programs and then double-checked by hand to guarantee the anomalies actually occur. 
\item We looked at the votes-by-round tables (tables of the form provided in Table \ref{first_STV_example}) for all 1,079 elections and attempted to find anomalies by hand for elections in which the vote totals in one of the rounds suggested that an anomaly might be present. We were unable to find any anomalous elections in this tedious, manual fashion beyond what our code found.
\item Our programs processed the elections over the course of several weeks across a few computers, and thus the computation time dedicated to searching for anomalies is quite substantial.
\item Similar programs have been used to find anomalies in single-winner ranked choice voting, and no anomalous elections have been found beyond those discovered by the programs \cite{GZ}, \cite{GM}.
\end{itemize}
Based on the steps described above, we believe that we have found all, or almost all, of the  Scottish STV elections which demonstrate an anomaly.

\section{Results}\label{results}

Of the 1,079 elections in the database we found a monotonicity anomaly of some type in 62 of them, 61  multiwinner and one single-winner. Table \ref{election_summary} summarizes our findings, providing a list of all elections which contain an anomaly and indicating which anomalies we are able to find in each election. Complete details reagrding each anomaly are available in the Appendix. Recall that these elections are  \emph{partisan}, meaning that each candidate runs as a member of a political party or runs as an independent, and thus we also provide information about when an anomaly affects a political party.  We use the following acronyms for the Scottish political parties: Conservative (Con), Green (Grn), Independent (Ind), Labour (Lab), Liberal Democrats (LD), and Scottish National Party (SNP).

\subsection{Committee Size Monotonicity Anomalies}

There are nine elections which demonstrate a committee size monotonicity anomaly, accounting for only $(9/1049)=0.86\%$ of the multiwinner elections in the database. Since we can definitively check for instances of this anomaly for a given election, we conclude that such anomalies should occur very infrequently in practice. 

While nine is a small sample size, these elections lead to several observations about committee size monotonicity anomalies in actual elections. First, a political party is harmed by this anomaly in only four elections. For example, in the 2012 Dundee City Ward 5 election the candidate McIrvine of the Labour Party loses their seat in the increase from $S=2$ to $S=3$, but the Labour Party receives exactly one seat for both values of $S$, and thus from the party's perspective it seems no harm was done. By contrast, in the 2017 East Dunbartonshire Ward 4 election Labour receives one seat when $S=3$ but receives zero seats in the actual election when $S=4$. From the perspective of political parties the rate of committee size anomalies is  $4/1049= 0.38\%$, suggesting that this anomaly should not be of concern to parties in real-world elections.

Second, in theory these anomalies can be quite extreme, in the sense that if an election contains enough candidates then it is possible that $W(P,S-1)$ and $W(P,S)$ are not only different, but also disjoint. We do not see such outlandish outcomes in the actual data, although we did find one election (2017 Moray Ward 3) where the IRV winner is not a member of the winner set $W(P,3)$. Our findings suggest that in real-world elections, when this anomaly occurs a single candidate loses their seat when $S-1$ seats is increased to $S$ seats.

Third, our code did not find any other type of anomaly in these nine elections. Thus our hypothetical example from Section \ref{preliminaries} which demonstrates all four anomaly types  represents a purely theoretical possibility. 

\subsection{Upward Monotonicity Anomalies}

We found 23 elections which demonstrate an upward monotonicity anomaly, accounting for $23/1079 = 2.1\%$ of the elections in the database. Twenty-two of the elections are multiwinner, providing a rate of $22/1049=2.1\%$ for elections with $S\ge 2$, and only one of the elections is single-winner, providing a rate of $1/30 =3.3\%$ for IRV elections. 

When an election contains an upward anomaly, it is perhaps not clear that harm has been done to any particular candidate. The winning candidate $X$ who would lose were they to be moved up on some ballots certainly isn't harmed, as the anomaly benefits them in a paradoxical way. It seems that if any candidate is harmed, it is a losing candidate $Y$ who would have won a seat if they had campaigned for $X$, causing $X$ to rise on some ballots and subsequently lose their seat in the resulting modified preference profile $P'$. We choose to say that such a candidate $Y$ is harmed by an upward anomaly, and if a political party wins more seats in the modified election $(P',S)$ than in the original election $(P,S)$, we say that this party has been harmed.

We found fifteen elections in which a political party was harmed by an upward anomaly. For example, in the 2022 Highland Ward 13 election, if MacKintosh of the Green Party were ranked higher on some ballots then Fraser of the Labour Party would replace MacKintosh in the winner set, suggesting that Labour should have done some carefully targeted campaigning for the Green Party. None of the examples found were as extreme as the hypothetical example from Section \ref{preliminaries}. In that example, if 6 voters who cast the ballot $D \succ A \succ C$ swapped $A$ and $D$ at the top of their ballots, then these voters would have caused both $A$ and $D$ to lose their seats, perhaps causing a party to lose two seats. We were unable to find any anomalies in the data where a set of voters would have caused their top $K\ge 2$ favorite candidates to lose their seats if those candidates were rearranged at the top of the voters' ballots.

We note that a monotonicity anomaly can sometimes illustrate just how ``close'' an election is. For the 2012 Aberdeenshire Ward 18 contest, in the original election candidate Samways received the fewest first place votes and was eliminated in the second of nine rounds. However, if the winning candidate Christie were moved up on some ballots, then Christie would eventually lose a seat and be replaced by Samways in the winner set. It seems odd that a candidate seemingly as weak as Samways could end up winning a seat through an upward anomaly, which we interpret as a sign of this election's competitiveness. 

Of the 23 elections demonstrating an upward anomaly, 19 also demonstrate a no-show anomaly and five 
 also demonstrate a downward anomaly.  For only three of the 23 elections could we not find some other type of monotonicity anomaly. While 23 is a small sample size, this suggests that upward anomalies tend to occur in conjunction with other anomalies in real-world STV elections.

\subsection{Downward Monotonicity Anomalies}

We found seventeen elections which demonstrate a downward monotonicity anomaly, nine strong and eight weak. All of these anomalies occur in multiwinner elections, and thus we obtain a rate of $17/1049=1.6\%$ for downward anomalies when $S \ge 2$, which drops to $8/1049=0.9\%$ for  strong anomalies.  One election demonstrates only downward and upward anomalies, two elections demonstrate only downward and no-show anomalies, and four elections demonstrate upward, downward, and no-show anomalies. We could not find any other kind of anomaly in the other ten elections which exhibit a downward anomaly.

In an election with a downward anomaly, it is clear which candidate and party (if any) have been harmed: if a candidate could have won a seat by being moved down on some ballots then this candidate is harmed by the anomaly, and if a party could have gained seats by having one of their candidates moved down on some ballots then the party is harmed as well. Of the seventeen elections demonstrating downward anomalies, a political party was harmed in fifteen of them. The Conservative Party seems to be the most affected by downward anomalies, with that party not winning a seat in seven of the fifteen elections as a result of this anomaly. For example, in the 2017 Argyll and Bute Ward 8 election, the Conservative Party did not win a seat in the original election but would have won a seat if their candidate Wallace were shifted down on some ballots.

As with the upward anomalies, none of the documented downward anomalies are as extreme as the hypothetical example from Section \ref{preliminaries}. We could not find any elections in which there exists a set of voters whose ballots start with $A\succ B$ and both $A$ and $B$ do not win a seat, but if $A$ were moved down on these ballots then both $A$ and $B$ win a seat. However, a few of the strong downward anomalies occur in a fashion we have not observed before. In a ``typical'' downward anomaly from prior literature, a losing candidate $A$ loses in the penultimate round to another candidate $B$, but when $A$ is shifted down on some ballots then $A$ is able to win by changing the elimination order so that $A$ no longer faces $B$ in that penultimate round. Our results show that downward anomalies in multiwinner elections can exhibit much different dynamics. For example, in the 2022 Perth and Kinross Ward 4 election Murray loses to Williamson by approximately 13.4 votes in the penultimate round, as shown in Table \ref{downward_example}.  If we shift Murray down one ranking on 37 ballots of the form Murray $\succ$ McDougall  then Murray still faces Williamson in the penultimate round but now Murray beats Williamson by approximately 7.74 votes. This anomaly occurs by swapping McDougall and Metcalf in the elimination order, but otherwise the order of elimination and election remains the same. It is strange that eliminating McDougall in the fourth round and eliminating Metcalf  in the sixth round results in Williamson winning a seat, but eliminating McDougall in the sixth round and eliminating Metcalf  in the fourth round results in Murray winning a seat. Some other examples of downward anomalies in our data are similarly strange when compared to downward anomalies from prior literature.

We do not have any insight into why strong downward anomalies  occur with much lower frequency than upward anomalies in the Scottish data. This empirical finding is consonant with prior theoretical work such as \cite{LDB} and \cite{Mi}, which show that upward anomalies occur much more frequently in IRV elections than strong downward anomalies\footnote{We note that \cite{LDB} and \cite{Mi}  use the term``downward monotonicity,'' which is equivalent to our notion of strong downward monotonicity.}. The most natural explanation for some of the discrepancy between downward and upward is that since voters tend to truncate their ballots, there is less opportunity for shifting a candidate down the rankings than if voters cast complete ballots (recall that for upward anomalies we can take a candidate which does not appear on the ballot and place them in the first ranking, as this does not change the relative rankings of the other candidates; thus, ballot truncation is not as much of an issue for the upward case). We do not have access to full rankings and thus are unable to substantiate this hypothesis.

\begin{table}

\begin{tabular}{c | c | c|c|c|c|c|c|c}
\multicolumn{9}{c}{\textbf{Actual Election}}\\
\hline
Candidate & \multicolumn{8}{c}{Votes by Round}\\
\hline
Duff (Con)& 1110 & \textbf{1120} &&&&&&\\
Hunter (Lab)& 147 & 166 & 166.18 &&&&&\\
McDade (Ind)& 977 & 1009 & 1010.10 & 1076.15 & \textbf{1148.16} &&&\\
McDougall (Grn)&203&212&212.10&247.11&&&&\\
McMahon (LD)&87&&&&&&&\\
Metcalf (Con) &268 & 275& 279.00 & 283.03 & 291.06 & 297.86 &&\\
Murray (SNP)&807&811&811.15&829.16&899.17&905.09 & 916.98&\\
Williamson (SNP)&856&857&857.16&865.16&908.16&914.89&930.38&\textbf{1740.42}\\
\hline

\end{tabular}

\vspace{.1 in}
\begin{tabular}{c | c | c|c|c|c|c|c|c}
\multicolumn{9}{c}{\textbf{Modified Election}}\\
\hline
Candidate & \multicolumn{8}{c}{Votes by Round}\\
\hline
Duff (Con)& 1110 & \textbf{1120} &&&&&&\\
Hunter (Lab)& 147 & 166 & 166.18 &&&&&\\
McDade (Ind) & 977 & 1009 & 1010.10 & 1076.15 & \textbf{1208.65} &&&\\
McDougall (Grn)&240&249&249.10&284.11&295.21&312.23&&\\
McMahon (LD)&87&&&&&&&\\
Metcalf (Con) &268 & 275& 279.00 & 283.03 &  &  &&\\
Murray (SNP)&770&774&774.15&792.16&795.21&806.18 & 950.61&\textbf{1759.20}\\
Williamson (SNP)&856&857&857.16&865.16&870.21&886.90&942.87&\\
\hline

\end{tabular}

\caption{The strong downward monotonicity anomaly in the 2022 election in the Highland Ward of the Perth and Kinross council area. The top table is constructed from the actual preference profile, and the bottom table is constructed from a modified profile in which Murray is shifted down on some ballots.}
\label{downward_example}

\end{table}

\subsection{No-show Anomalies}
We found 39 elections which demonstrate a no-show anomaly, accounting for $39/1079 = 3.6\%$ of the elections in the database, and a political party was harmed in 29 of them. The Labour Party is the most affected by this anomaly, with twelve of the 29 elections featuring a losing Labour candidate who would have won a seat if some of their supporters abstained (and this candidate would not simply replace another Labour candidate in the winner set). Thirty-eight of the 39 elections are multiwinner; we found a no-show anomaly in only one of the single-winner elections. Nineteen of the elections also demonstrate an upward anomaly and six 
also demonstrate a downward anomaly. 

For 36 of the 38 multiwinner elections demonstrating a no-show anomaly we could not find a set of ballots to remove such that the affected candidate is ranked in the top $S$ rankings on all removed ballots. These elections are marked with a $\dagger$ in Table \ref{election_summary}. For example, in the 2022 Fife Ward 10 election if we remove 93 ballots on which the losing candidate Smart is ranked above the winning candidate Leslie then Smart replaces Leslie in the winning set, but for some of these ballots Smart is not ranked in the voters' top four candidates.

\begin{table}

\begin{tabular}{c | c | c|c|c|c}
\multicolumn{6}{c}{\textbf{Actual Election}, Quota$=$299}\\
\hline
\hline
Candidate & \multicolumn{5}{c}{Votes by Round}\\
\hline
Campbell & \textbf{299} &&&&\\
Smith & \textbf{429} &&&&\\
Valente & 180 & 220.61 & 268.35&268.35 & \\
Westlake & 190 & 242.42 & 271.12 & 271.12&\textbf{408.83}\\
Wishart & \textbf{394} &&&&\\
\hline

\end{tabular}

\vspace{.1 in}

\begin{tabular}{c | c | c|c|c|c}
\multicolumn{6}{c}{\textbf{Modified Election}, Quota$=$299}\\
\hline
\hline
Candidate & \multicolumn{5}{c}{Votes by Round}\\
\hline
Campbell  & 298 &\textbf{364.36}&&&\\
Smith & \textbf{429} &&&&\\
Valente & 180 & 200.30 & 248.04 &270.34& \textbf{420.64}\\
Westlake & 190 & 215.76 & 244.22 & 270.22&\\
Wishart & \textbf{394} &&&&\\
\hline

\end{tabular}

\caption{(Top) The votes-by-round table for the 2017 council election in the seventh ward of the Shetland Islands council area. (Bottom) A votes-by-round table demonstrating a no-show anomaly in this election.}
\label{no-show-ex}

\end{table}

We found eighteen elections which demonstrate a no-show anomaly but do not demonstrate any other type which could be found by our code (or which could be found by manual inspection). An example is the 2017 election in the seventh ward of the Shetland Islands, the votes-by-round table for which is displayed in the top of Table \ref{no-show-ex}. Note that Valente does not win a seat while Westlake does. If we remove one ballot of the form Campbell $\succ$ Valente  $\succ$ Smith  $\succ$ Wishart  $\succ$ Westlake then the election plays out as shown in the bottom of Table \ref{no-show-ex}. By removing this ballot we delay the election of Campbell by one round, which eventually allows Valente to win narrowly over Westlake. Prior to our analysis of the Scottish database, there were no known real-world elections demonstrating only a no-show anomaly (based on the ballot data for the Shetland election, it is straightforward to show that we cannot find any other anomaly in this election.) 

\newpage
\begin{longtable}{l|c|c|c|c|c}

Election & $S$ & Comm. Size&Upward & Downward & No-show\\
\hline
2012 Dundee Ward 5 & 3 & \textbf{Yes} & No & No & No\\
2012 N-Lanarks Ward 3 & 4& \textbf{Yes} & No & No & No\\
2017 Dumgal Ward 12 & 3 & \textbf{Yes} & No & No & No\\
2017 E-Duns Ward 4 & 4& \textbf{Yes} & No & No & No\\
2017 Moray Ward 3 & 4& \textbf{Yes} & No & No & No\\
2017 W-Duns Ward 3 & 4& \textbf{Yes} & No & No & No\\
2022 E-Duns Ward 4 & 4& \textbf{Yes} & No & No & No\\
2022 Edinburgh Ward 15 & 4& \textbf{Yes} & No & No & No\\
2022 S-Lanarks Ward 9 & 3& \textbf{Yes} & No & No & No\\
2012 Aberdeenshire Ward 18 & 4&No & \textbf{Yes} & No & \textbf{Yes}\\
2012 Argyll Bute Ward 5 & 4&  No &\textbf{Yes} & No & \textbf{Yes}\\
2012 Eilean Siar Ward 5 & 3 &  No &\textbf{Yes} & No & \textbf{Yes}\\
2012 Eilean Siar Ward 7 & 4 &  No &\textbf{Yes} & \textbf{Yes (S)} & \textbf{Yes}\\
2012 Highland Ward 7 & 4 & No & \textbf{Yes} &\textbf{Yes (S)} &\textbf{Yes}\\
2012 Highland Ward 20 & 4& No & \textbf{Yes} &\textbf{Yes (S)} &\textbf{Yes}\\
2012 Perth-Kinross Ward 9 & 3 & No & \textbf{Yes} & No &\textbf{Yes}\\
2017 Argyll Bute Ward 8 & 3&No & \textbf{Yes} &\textbf{Yes (S)} &No\\
2017 E-Duns Ward 6 & 3 & No & \textbf{Yes} &No &No\\
2017 Edinburgh Ward 4 & 4 &No & \textbf{Yes} & No &\textbf{Yes}\\
2017 Fife Ward 12 & 3 & No & \textbf{Yes} &No &\textbf{Yes}\\
2017 Glasgow Ward 5 & 4 & No & \textbf{Yes} &No & No\\
2017 Glasgow Ward 9 & 4 & No & \textbf{Yes} &No &\textbf{Yes}$^\dagger$\\
2017 N-Lanarks Ward 3 & 4 & No & \textbf{Yes} &No &\textbf{Yes}\\
2017 Perth-Kinross Ward 10$^*$ & 1 & No & \textbf{Yes} & No & \textbf{Yes}$^\dagger$\\
2022 Aberdeenshire Ward 18 & 4 & No & \textbf{Yes} & No & \textbf{Yes}\\
2022 Dumgal Ward 7&3 &No & \textbf{Yes}&  No& \textbf{Yes} \\
2022 Edinburgh Ward 5 & 4 & No & \textbf{Yes} & \textbf{Yes (S)} & \textbf{Yes}\\
2022 Fife Ward 10 & 3 &No & \textbf{Yes}&  No& \textbf{Yes}$^\dagger$ \\
2022 Glasgow Ward 13 &4 &No & \textbf{Yes} &  No& \textbf{Yes} \\
2022 Highland Ward 13 & 3 & No & \textbf{Yes} & No & \textbf{Yes}\\
2022 Orkney Ward 5 & 3 & No & \textbf{Yes} & No & No\\
2022 S-Lanarks Ward 12 &3&No & \textbf{Yes}&  No& \textbf{Yes} \\
2017 Aberdeenshire Ward 9&4 & No & No& \textbf{Yes  (S)} & \textbf{Yes} \\
2017 N-Ayrshire Ward 9 & 3 &No & No& \textbf{Yes  (W)} & No \\
2017 N-Lanarks Ward 16 &3 & No & No& \textbf{Yes  (S)} & No \\
2017 N-Lanarks Ward 20 &4 & No & No& \textbf{Yes  (W)} & No \\
2017 Stirling Ward 3& 4 & No & No& \textbf{Yes  (W)} & No \\
2017 Renfrewshire Ward 6 & 3& No & No& \textbf{Yes  (W)} & No \\
2022 Aberdeen City Ward 8 & 4 & No & No& \textbf{Yes  (W)} & No \\
2022 Aberdeenshire Ward 8&4&No & No& \textbf{Yes  (W)} & No \\
2022 Argyll Bute Ward 8&3&No & No& \textbf{Yes  (W)} & \textbf{Yes} \\
2022 Falkirk Ward 2 & 3 &No & No& \textbf{Yes (S)} & No \\
2022 Glasgow Ward 23 & 4 & No & No & \textbf{Yes  (W)}  & No\\
2022 Perth-Kinross Ward 4 & 3 &  No & No & \textbf{Yes  (S)}  & No\\
2012 E-Ayrshire Ward 5 &3 &No & No & No & \textbf{Yes} \\
2012 Falkirk Ward 1 & 3 &No & No & No & \textbf{Yes} \\
2012 Highland Ward 2 & 3 &No & No & No & \textbf{Yes} \\
2012 Highland Ward 9&4 &No & No & No & \textbf{Yes} \\
2012 Moray Ward 5 & 4 &No & No & No & \textbf{Yes} \\
2012 Moray Ward 6 & 3 &No & No & No & \textbf{Yes} \\
2017 Dumgal Ward 9 & 4 & No & No & No & \textbf{Yes} \\
2017 E-Ayrshire Ward 6 & 3 &No & No & No & \textbf{Yes} \\
2017 Edinburgh Ward 8 & 3&No & No & No & \textbf{Yes} \\
2017 Moray Ward 5 & 4 &No & No & No & \textbf{Yes} \\
2017 N-Lanarks Ward 11 &4 &No & No & No & \textbf{Yes} \\
2017 Shetland Ward 7 & 4 & No & No & No & \textbf{Yes} \\
2017 W-Duns Ward 2 & 4 & No & No & No & \textbf{Yes} \\
2022 Argyll Bute Ward 4 & 4 & No & No & No & \textbf{Yes} \\
2022 Edinburgh Ward 2 & 4 & No & No & No & \textbf{Yes} \\
2022 Fife Ward 7 & 4 &No & No & No & \textbf{Yes} \\
2022 N-Lanarks Ward 17 & 3 & No & No & No & \textbf{Yes} \\
2022 W-Lothian Ward 7 & 4 & No & No & No & \textbf{Yes} \\

\caption{ The one single-winner (out of 30) and 61 multiwinner (out of 1049) elections which demonstrate an anomaly. The last four columns denote the four types of monotonicity anomaly. The S (resp. W) in the Downward column denotes that the downward anomaly is strong (resp. weak). The * denotes this was a by-election. The $\dagger$ denotes this no-show anomaly is weak in the sense that  we could not find a set of ballots where the affected candidate is ranked in the top $S$ candidates.}
\label{election_summary}

\end{longtable}

\section{Discussion: Multiwinner Versus Single-Winner Anomalies}

Monotonicity anomalies (besides committee size anomalies) in the single-winner case have been well-studied for elections with a small number of candidates. Prior research has shown that for a single-winner election to demonstrate an anomaly, the following conditions must be met.

\begin{itemize}
\item The election contains at least three ``viable'' candidates. For example, as discussed in\cite{Mi} and \cite{ON},  for an upward anomaly to occur in a 3-candidate election each candidate must earn at least 25\% (and less than 50\%) of the first-place votes.
\item The anomaly occurs by changing ballots in such a way that the order of elimination of candidates changes, resulting in different candidates surviving to subsequent rounds.
\end{itemize}

The single-winner 2017 by-election from the Perth City South Ward in the Perth and Kinross council area provides a typical example of an upward anomaly. The election contained six candidates; the preference profile in Table \ref{perth-kinross-ex} shows the ballot data after the bottom three candidates were eliminated, leaving surviving candidates Barrett (B), Coates (C), and Leitch (L). For convenience, we combine ballots of the form $C_1 \succ C_2 \succ C_3$ with ballots of the form $C_1 \succ C_2$, as each type of ballot conveys the same ranking information.
Table \ref{perth-kinross-ex} shows that the first-place vote totals are 1733, 1762, and 1883 for B, C, and L, respectively. Thus Barrett is eliminated and, after the vote transfers, Coates wins with 2381 votes to Leitch's 2227. If we change 151 bullet votes for Leitch to ballots of the form Coates $\succ$ Leitch, then Coates is no longer the winner of the resulting election. The reason is that in the modified election Leitch (instead of Barrett) is eliminated first, and the vote totals in Table \ref{perth-kinross-ex} show that Coates would then lose head-to-head against Barrett.

This election exhibits a no-show anomaly in a similar fashion. If we remove 151 ballots of the form Leitch $\succ$ Barrett $\succ$ Coates then in the resulting election Leitch is eliminated first and Barrett still has enough support to defeat Coates head-to-head in the final round. The 151 voters who cast these ballots would have achieved a more desirable electoral outcome (obtaining their second choice instead of their third) if they had not participated in the election.
\begin{table}

\begin{tabular}{l|c|c|c|c|c|c|c|c|c}
Num. Voters & 770 & 619 & 344 & 846 & 867 & 49 & 1167 & 620 & 96\\
\hline
1st choice & $B$ & $B$ &$B$ & $C$ & $C$ & $C$ & $L$ & $L$ & $L$ \\
2nd choice & & $C$ & $L$ & & $B$ & $L$ && $B$ & $C$\\
3rd choice & & $L$ & $C$ & & $L$ & $B$ & & $C$ & $B$\\
\end{tabular}

\caption{The preference profile for the 2017 by-election in the tenth ward of the Perth and Kinross council area, after eliminating the bottom three candidates.}
\label{perth-kinross-ex}

\end{table}

Note that this election contains the hallmark features of an IRV election demonstrating an  anomaly as outlined above. All three candidates are ``strong'' in the sense that their first-place votes totals surpass 25\% of the total votes, and we engineer an anomaly by changing the order of elimination so that the set of candidates surviving in the last round is different from that in the original election. All other documented examples of real-world IRV elections demonstrating upward, downward, or no-show anomalies display a similar dynamic \cite{GM}.

This example (and prior research) raises the questions: in the multiwinner case, must the election contain $S+2$ ``viable'' or ``strong'' candidates? Must an anomaly occur by changing which candidates survive to a subsequent round? Interestingly, the answer to both questions is No. To show why, consider the example of the 2017 election in the ninth ward of the Aberdeenshire council area, the  votes-by-round table for which is shown in the top of Table \ref{aberdeenshire-2017-09}. Note that all candidates except Morgan win a seat, and there are not $S+2$ viable candidates. If we shift Morgan down to the second ranking on 17 ballots on which Morgan is ranked first and Davidson is ranked second, the election unfolds as shown in the bottom table of Table \ref{aberdeenshire-2017-09}. If Morgan gives 17 of his first-place votes to Davidson as described then Davidson makes quota in the first round and thus is ineligible to receive vote transfers from the election of Owen, and as a result Morgan wins a seat instead of Kahanov-Kloppert. That is, we engineer an anomaly by making Davidson achieve quota a round earlier. The same three candidates as in the original election survive to the penultimate round, and yet the final two candidates to receive a seat are different. This Aberdeenshire example shows that monotonicity anomalies in multiwinner elections can arise in surprising and interesting ways that would not be anticipated from the study of single-winner STV elections.

\begin{table}

\begin{tabular}{c | c | c|c|c}
\multicolumn{5}{c}{\textbf{Actual Election}, Quota$=$1104}\\
\hline
\hline
Candidate & \multicolumn{4}{c}{Votes by Round}\\
\hline
Davidson & 1087 & \textbf{1703.86} & &\\
Kahanov-Kloppert & 768 & 739.55 & 843.88 & \textbf{1024.70}\\
Morgan  & 415 & 548.39 & 835.99 &\\
Owen & \textbf{2258} &&&\\
Thomson & 987 & 1034.02 & 1093.74 & \textbf{1264.34}\\
\hline

\end{tabular}

\vspace{.1 in}

\begin{tabular}{c | c | c|c|c}
\multicolumn{5}{c}{\textbf{Modified Election}, Quota$=$1104}\\
\hline
\hline
Candidate & \multicolumn{4}{c}{Votes by Round}\\
\hline
Davidson & \textbf{1104} & & &\\
Kahanov-Kloppert& 768 & 814.51 & 814.51 &\\
Morgan  & 398 & 851.32 & 851.32 &\textbf{922.07}\\
Owen & \textbf{2258} &&&\\
Thomson & 987 & 1063.67& 1063.67 & \textbf{1743.66}\\
\hline

\end{tabular}
\caption{(Top) The votes-by-round table for the 2017 council election in the ninth ward of the Aberdeenshire council area. (Bottom) A votes-by-round table demonstrating a downward anomaly in this election.}
\label{aberdeenshire-2017-09}

\end{table}

Many of the elections demonstrating no-show anomalies, particularly the elections which do not demonstrate any other type of anomaly, display dynamics that are even further from the single-winner case than the Aberdeenshire election discussed above. In the Aberdeenshire example,  we create a downward anomaly by changing the round in which Davidson achieves quota. For many of the no-show anomalies, we can engineer the anomaly without changing the round in which any candidate is elected or eliminated, until the round in which a different candidate wins a seat. To see an example, consider the 2022 council election in the seventh ward of the West Lothian council area, the votes-by-round table for which is displayed in the top of Table \ref{w-lothian-2017-07}. Note that Paul surpasses quota by 9 votes, 7.08 of which are transferred to fellow Labour candidate Sullivan and 0.04 of which are transferred to Fairbairn. As these numbers suggest, Sullivan is the candidate who benefits most from the election of Paul. The additional 7.08 votes eventually allow Sullivan to win the last seat, as Sullivan defeats Fairbairn by approximately two votes.

  If we remove 10 ballots of the form Paul $\succ$ Fairbairn then the election unfolds as shown in the bottom table of Table \ref{w-lothian-2017-07}. Note that the removal of these ballots causes Paul to exceed quota by only one vote, dramatically decreasing the number of votes transferred to Sullivan (Fairbairn, on the other hand, barely feels the difference). As a result, Fairbairn defeats Sullivan by approximately four votes in the penultimate round and replaces Sullivan in the winner set. 

The modified election which demonstrates the no-show anomaly unfolds in exactly the same order as the original election; there is no change in the order or elimination of candidates until the penultimate round, when Sullivan is eliminated instead of Fairbairn. This kind of phenomenon is not possible to observe in the single-winner case, where we can engineer a no-show anomaly only by changing the order of elimination of the candidates. This example illustrates that the multiwinner setting can provide a much richer set of no-show anomaly examples.

\begin{table}

\begin{tabular}{c | c | c|c|c|c|c}
\multicolumn{7}{c}{\textbf{Actual Election}, Quota$=$1228}\\
\hline
\hline
Candidate & \multicolumn{6}{c}{Votes by Round}\\
\hline
J. Dickson (SNP)& \textbf{1734} &&&&&\\
M.Dickson (SNP)& 636 &1085.10 & 1085.31 & 1095.50 & 1195.55 & \textbf{1230.47}\\
Fairbairn (Con)& 1135 & 1138.50 & 1138.99 & 1175.03 & 1187.34 & \\
Pattle (LD)& 148 & 152.09 & 152.36 &&&\\
Paul (Lab) & \textbf{1237} &&&&&\\
Racionzer (Grn) & 190 & 203.42 & 203.62 & 231.13 &&\\
Sullivan (Lab) & 1056 & 1068.55 & 1075.63 & 1121.62 & 1189.20 & \textbf{1746.90}\\

\hline

\end{tabular}

\vspace{.1 in}

\begin{tabular}{c | c | c|c|c|c|c}
\multicolumn{7}{c}{\textbf{Modified Election}, Quota$=$1226}\\
\hline
\hline
Candidate & \multicolumn{6}{c}{Votes by Round}\\
\hline
J. Dickson (SNP)& \textbf{1734} &&&&&\\
M.Dickson (SNP)& 636 &1086.87 & 1086.89 & 1097.06 & 1197.15 & \textbf{1425.02}\\
Fairbairn (Con)& 1135 & 1138.52 & 1138.56 & 1174.57 & 1186.86 &\textbf{1476.65} \\
Pattle (LD)& 148 & 152.10 & 152.13 &&&\\
Paul (Lab) & \textbf{1227} &&&&&\\
Racionzer (Grn) & 190 & 203.48 & 203.50 & 203.97 &&\\
Sullivan (Lab) & 1056 & 1068.60 & 1069.39 & 1115.28 & 1182.76 &\\

\hline

\end{tabular}
\caption{(Top) The votes-by-round table for the 2022 council election in the seventh ward of the West Lothian council area. (Bottom) A votes-by-round table demonstrating a no-show anomaly in this election.}
\label{w-lothian-2017-07}
\end{table}


Perhaps the primary factor that distinguishes the multiwinner STV setting from the single-winner is that a multiwinner election often continues for many rounds after a candidate achieves quota. This creates opportunities for engineering anomalies by changing the ballot data so that a winning candidate achieves quota in a different round or surpasses quota by a different amount. Tables \ref{aberdeenshire-2017-09} and \ref{w-lothian-2017-07} both illustrate this point, and there are many other such examples. The use of a quota introduces a potentially ``distortionary'' effect into the vote transfer process, where if a candidate is close to earning quota or just barely surpasses quota in a given round then there is a a chance to create an anomaly. In the case of no-show anomalies, the use of a quota can create dynamics as observed in Table \ref{w-lothian-2017-07} where a minority group of voters who do not express support for a single party can pay a price for casting their votes.

The multiwinner setting also produces different, higher anomaly rates than in the single-winner case. The most comprehensive empirical study of anomaly rates for single-winner STV elections is \cite{GMb}, which found rates of 2.2\%, 1.5\%, and 0.5\% for upward, downward, and no-show anomalies, respectively, in a large database of American STV elections. However, that study used a database of elections in which no candidate earned an initial majority of first-place votes, which dramatically decreased the denominator used to calculate the reported rates. (In the single-winner case, if an election contains such a majority candidate then we know \emph{a priori} that the election does not demonstrate any anomaly.) If we were to include American single-winner STV elections in which a candidate wins in the first round, the reported rates would decrease significantly. Thus, multiwinner STV elections seem to produce much higher anomaly rates than single-winner elections. Perhaps this is to be expected: if there are more winning candidates then there are more opportunities for a candidate to be affected by an upward anomaly, for example. 

What is more surprising is that the rates of each anomaly type relative to the others is much different in the multiwinner case than what the single-winner setting would suggest. Of the five single-winner elections which demonstrate any type of anomaly in \cite{GMb}, four demonstrate an upward anomaly, three demonstrate a downward anomaly, and only one demonstrates a no-show anomaly. Five is a small sample size, but these findings suggest that upward anomalies occur at a higher rate than the other two, and no-show anomalies are incredibly rare (prior to the work in this article, the only documented no-show anomaly in a real-world election occurred in the August 2022 Special Election for US House in Alaska \cite{GM}). In the multiwinner setting, no-show anomalies seem to occur at a much higher rate than other anomalies. The reason seems to be that we can create examples such as the election in Table \ref{w-lothian-2017-07}, where an anomaly is exhibited without having to change the order of elimination or election of candidates in previous rounds.

\section{Discussion: Close Elections}\label{discussion}
In this section we discuss our results through an examination of how frequently anomalies arise in close multiwinner elections, since much of the prior literature focuses on the frequency of monotonicity anomalies in elections that are \emph{close} in some sense. For example, \cite{M} and  \cite{ON} examine the single-winner case with $n=3$, and they define an election to be close if all three candidates receive more than 25\% of the first-place votes. Both papers then argue that monotonicity anomalies are much more likely to occur in such close elections.

To build on this literature, we investigate how much closeness matters for monotonicity anomalies in the 1,049 multiwinner Scottish elections. The primary difficulty of such an investigation is that \emph{closeness} is more difficult to define in the multiwinner setting with more than three candidates. We briefly examine several reasonable notions of multiwinner closeness.

If all $S$ winners achieve quota in the first round, we know without examining the ballot data that it is not possible for the election to demonstrate an upward, downward, or no-show anomaly. Such elections are analogous to single-winner elections in which a candidate achieves a majority of votes in the first round, which is a common occurrence in other election databases such as municipal IRV elections in the United States.  Our first notion of closeness is that the election does not terminate after only one round, so that not all winners achieve quota initially. Of the 1,049  multiwinner elections in the database 1,026 satisfy this notion of closeness, and thus it is rare for a Scottish election to terminate in the first round. Using a denominator of 1,026 rather than 1,049 does not significantly alter the percentages provided in the previous section.

\begin{figure}[tbh] 
\begin{center}
\includegraphics[scale=0.8]{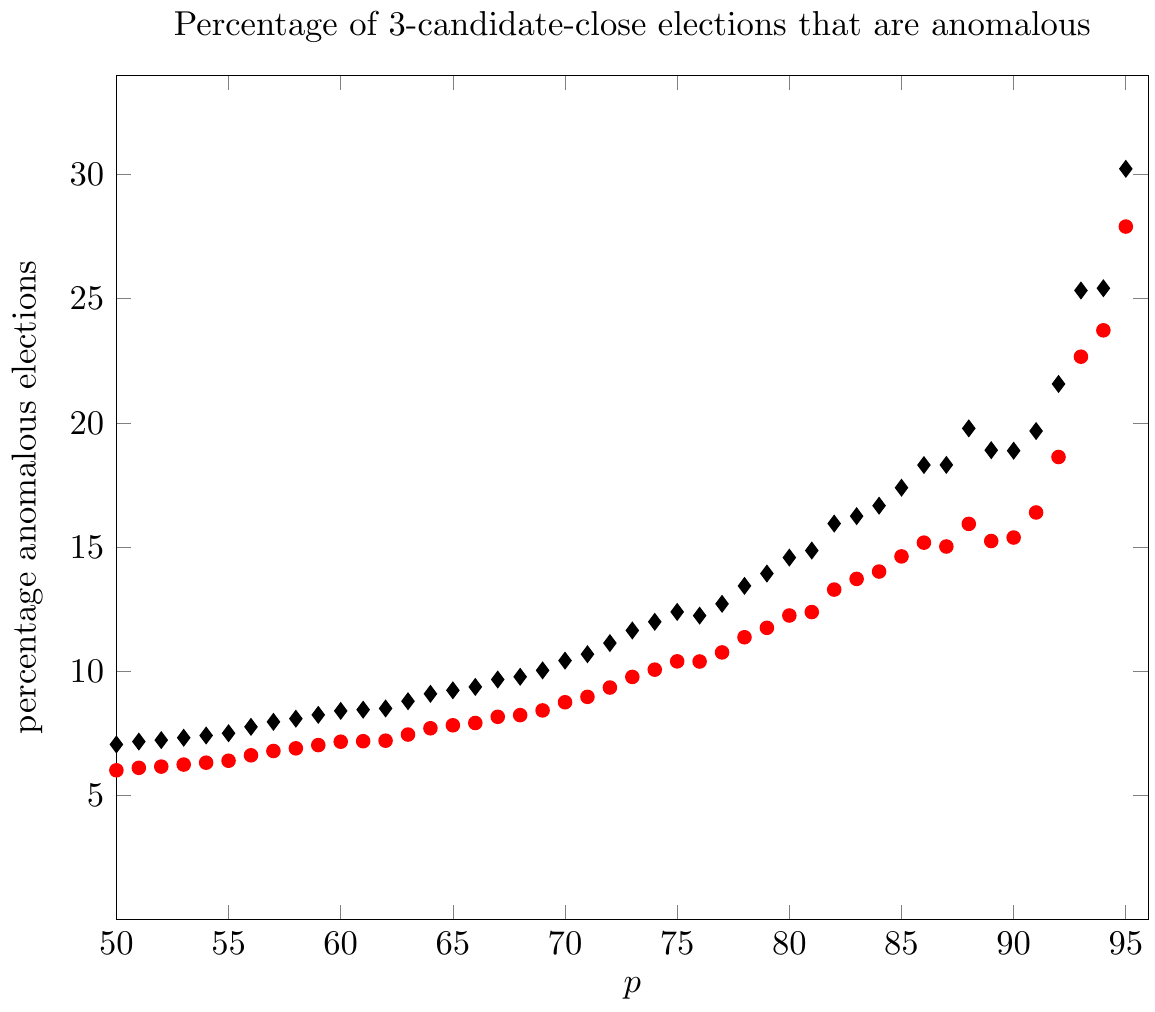}

\end{center}
\caption{For a given $p$, the percentage of  three-candidate-close elections that demonstrate an anomaly. The black diamonds use elections which demonstrate any anomaly, while the red disks exclude elections demonstrating a committee size anomaly.}
\label{closeness3cands}
\end{figure}

Next, we generalize the notion of closeness found in \cite{M}, which states that a three-candidate election is \emph{close} if the candidate with the fewest first-place votes has at least half as many first-place votes as the candidate with the most. Mimicking this idea, we say that an election is \emph{three-candidate-close} with parameter $p$ if there exists a round of the election and a three candidate subset of candidates who have not been eliminated or previously elected in this round such that (1) this subset of candidates contains at least one candidate who eventually wins a seat and one candidate who does not win a seat, and (2) the smallest of the vote totals for the three candidates in this round is at least $p$\% of the largest vote total. The black diamonds of Figure \ref{closeness3cands} show the percentage of three-candidate-close elections which are anomalous for $p \in \{50, 51, \dots, 95\}$. For example, when $p=50$ all 61 anomalous elections are three-candidate-close while 865 of the multiwinner elections in the database are, and thus we plot a black diamond at height $61/865 = 7.1\%$. Note as $p$ increases the percentage of close elections demonstrating an anomaly also generally increases. When $p=95$ there are 43 three-candidate-close multiwinner elections, 13 of which demonstrate an anomaly, yielding a percentage of 30.2\%. The red disks show similar information but exclude the elections demonstrating a committee size anomaly from the numerator.

\begin{figure}[tbh] 
\begin{center}
\includegraphics[scale=0.8]{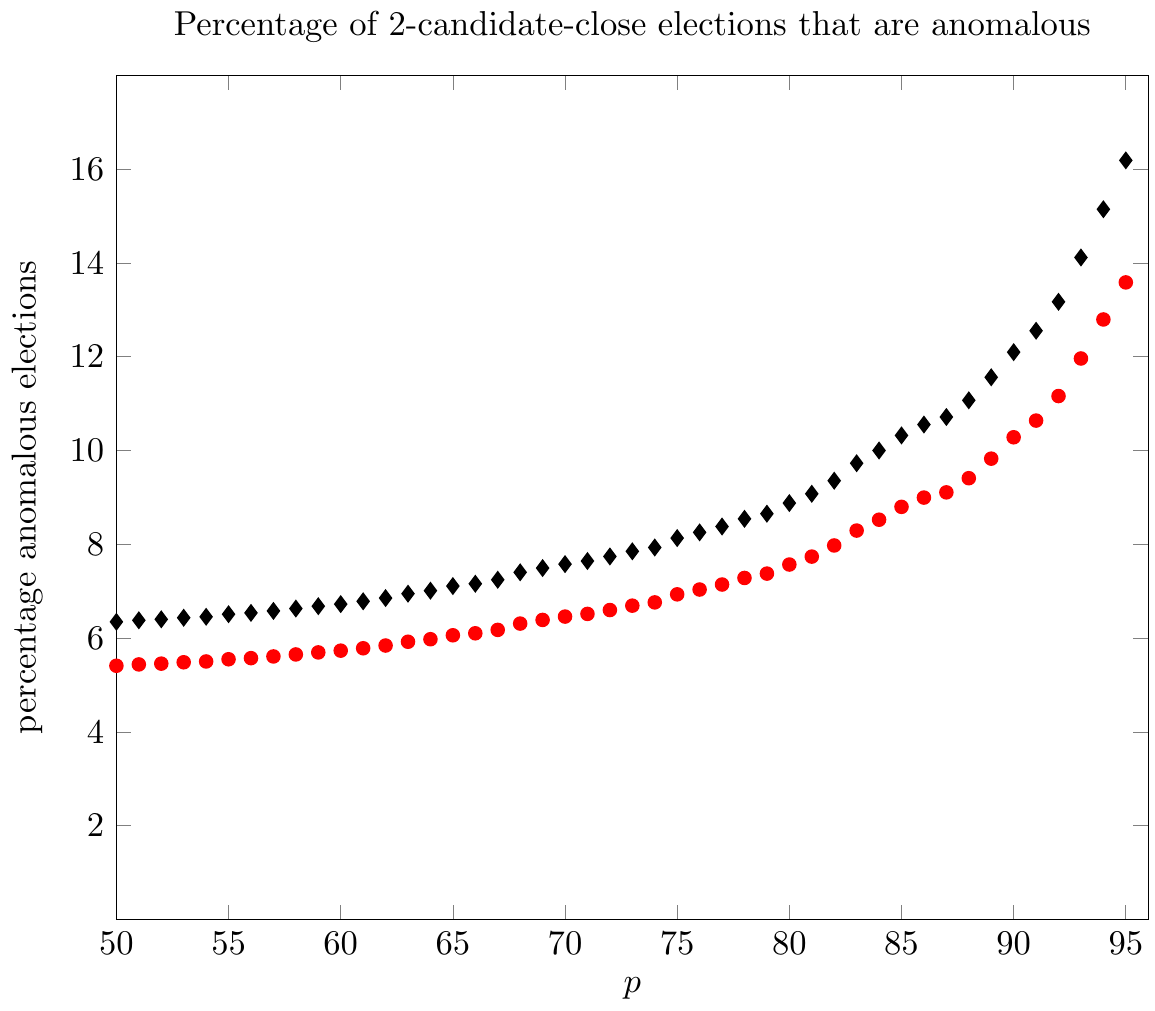}

\end{center}
\caption{For a given $p$, the percentage of  two-candidate-close elections that demonstrate an anomaly. The black diamonds use elections which demonstrate any anomaly, while the red disks exclude elections demonstrating a committee size anomaly.}
\label{closeness2cands}
\end{figure}

In the popular discourse around elections, often an election is called ``close'' if one candidate narrowly misses winning a seat in some sense. In an attempt to make this notion rigorous, we say that an election is \emph{two-candidate-close} with parameter $p$ if there exists a round of the election and a two candidate subset of candidates who have not been eliminated or previously elected in this round such that (1) one of the candidates eventually wins a seat and the other does not win a seat, and (2) the smaller of the vote totals of the two candidates in this round is at least $p$\% of the larger. The black diamonds of Figure \ref{closeness2cands} show the percentage of two-candidate-close elections which are anomalous for $p \in \{50, 51, \dots, 95\}$. As with three-candidate-closeness, as we increase $p$ the percentage of anomalous elections also increases, although we do not obtain percentages nearly as high as the three-candidate-closeness case. There are 346 elections which are two-candidate-close with $p=95$, fifty-six of which are anomalous, producing a maximal percentage of 16.2\%. The red disks show similar information but exclude the elections demonstrating a committee size anomaly from the numerator.

In the single-winner setting, often elections are described as ``close'' if a Condorcet winner does not exist or if different voting methods choose different winners. Building on this notion, we test if we obtain a higher percentage of anomalous elections if we restrict to elections which do not contain a set of candidates which generalize the notion of a Condorcet winner, or which produce different winner sets under different multiwinner voting rules.

To generalize the idea of a Condorcet winner, we use the concept of a \emph{Condorcet committee} as defined in  \cite{Gehr} and \cite{R2}. A set of $S$ candidates $\mathcal{C}$ is a \emph{Condorcet committee} if for each pair of candidates $(A,B)$ with $A \in \mathcal{C}$ and $B \not \in \mathcal{C}$, more voters prefer $A$ over $B$ than the reverse. When $S=1$, if a Condorcet committee exists then the single candidate in the set is the election's Condorcet winner. If such a candidate does not exist, the election can be viewed as especially close or competitive. Of the 1,049 multiwinner elections in the database only 18 do not contain a Condorcet committee of size $S$. Four of these elections demonstrate an anomaly, giving an anomaly rate of 22.2\% for these very close elections.

We can use the concept of a Condorcet committee to build voting methods that are the multiwinner analogue of the single-winner Condorcet methods. To be precise, we can create methods which select the Condorcet committee of size $S$ if one exists, and otherwise uses some kind of pairwise comparison methodology to select the winner set. Since most of our elections contain a Condorcet committee we do not choose among such Condorcet methods, but instead focus only on the elections with such a committee. We can then say that an election is not close if the winner set under other voting methods all equal the Condorcet committee. Consider the method of single non-transferable vote (SNTV), for example. Under this method, the winner set consists of the $S$ candidates with the most first-place votes. While STV aims to achieve proportional representation, SNTV aims to achieve some level of semi-proportional representation, and a Condorcet method does not attempt to achieve any level of proportional representation. If these three methods all choose the same winner set, in some sense the election is not close. 414 elections satisfy this condition, meaning that from this perspective 635 elections are close. Of these, 48 demonstrate some type of anomaly, yielding an anomaly rate of 7.6\% for elections which are close in the sense that different election methods would select a different winners set.
 
 There has been no prior theoretical work on closeness and the frequency of monotonicity anomalies for multiwinner STV elections, and thus we cannot directly compare our percentages to prior work.  However, there has been substantial research related to closeness for 3-candidate IRV elections. Our percentages are much lower than what is predicted by \cite{Mi} or \cite{ON}, both of which give probabilities between 12.5\% and 51\% for an election to demonstrate an upward or downward anomaly in closely contested single-winner contests, with the highest percentages found for the most competitive elections. Our work confirms prior observations that the closeness of an election matters for the frequency of monotonicity anomalies, and depending on the notion of closeness used we achieve percentages that are consonant with some of the prior theoretical work (although we never achieve anything close to 51\%). 
  
 It is unsurprising that the percentages we find are generally lower than what occurs under theoretical models, for two main reasons. Firstly, the theoretical models tend to provide upper bounds for the frequency of an anomaly occurring. That is, theoretical models often provide the ``worst-case'' scenario because these models can produce elections which contain conflicted electorates at a higher proportion that what we see in practice. For example, under the random impartial culture and impartial anonymous culture models, IRV has a much larger tendency not to choose a Condorcet winner than we observe in actual elections (see \cite{GL} for a summary of the theoretical results, and \cite{MM2} and \cite{Song} for an empirical analysis).   Secondly, as noted previously, there is a very high rate of ballot truncation in the Scottish elections, which likely reduces the frequency of anomalies as compared to theoretical work which uses exclusively full ballots.  It is unknown precisely what affect ballot truncation has on anomaly rates, however, which is an area for further study.  For these reasons, our lower percentages than the theoretical work is entirely expected.

\section{Conclusion}\label{conclusion}

The 62 elections demonstrating monotonicity anomalies that we found, including the 53 elections which contain an upward, downward, or no-show anomaly, seem to undermine the claims of \cite{A}, \cite{B}, and \cite{G}, which essentially state that monotonicity anomalies either do not occur in actual STV elections or occur extremely rarely and therefore monotonicity issues are of no practical concern. On the other hand, the anomaly rates we found are not particularly large, unless we restrict to small sets of extremely competitive elections. Essentially, our findings suggest that an anomaly of each type should occur about 3-13 times on average per election cycle, which is small but not minuscule compared to the approximately 350 contested STV elections which occur across Scotland in a local government election year.  We remind the reader that we cannot guarantee that we found all anomalous elections and thus more sophisticated code could potentially find more anomalies, perhaps bringing the anomaly rate more in line with estimates from the single-winner literature. The problem of deciding whether a given preference profile demonstrates a particular anomaly (besides a committee size anomaly) in an STV election is computationally quite difficult, and is an interesting avenue for future work. 

We note that for some researchers monotonicity anomalies are problematic because they leave open the possibility that voters could strategically manipulate the election outcome by voting insincerely. Based on our results, it seems highly unlikely that deliberate strategic voting to engineer a monotonicity anomaly is a problem in practice. The complex nature of the STV algorithm for multiwinner elections seems to preclude this possibility, and voters would need access to preference poll data of extremely high quality to create such an outcome.  In our view, monotonicity anomalies in the multiwinner setting are offensive only in hindsight, when it becomes apparent after an election that a candidate, a party, or some voters could have obtained a better electoral outcome in a paradoxical fashion. Furthermore, the number of voters used to exhibit an anomaly is small and the set of voters must be precisely chosen, and thus the number of voters  negatively affected by these anomalies is much lower than what might be suggested by the reported anomaly rates. For example, even though the percentage of elections which demonstrate a no-show anomaly is 3.6\%, the number of ballots we can remove to create the anomalies represents a tiny fraction of the total number of voters across these elections. It is not the case that 3.6\% of all voters would create better electoral outcomes by not voting.

What does the presence of these anomalies in the Scottish elections say about the use of STV? Does STV's susceptibility to these anomalies in actual elections imply that STV should not be used? These questions cannot be answered mathematically, as the answers depends on value judgements outside mathematics. If we take the reasonable position that monotonicity anomalies are offensive enough that methods susceptible to such outcomes should be discarded, then this article is a strong argument against the use of STV. On the other hand, if we feel that STV has benefits which outweigh the low rate of monotonicity anomalies we found in the Scottish data, then this article does not undermine the use of STV. Either way, we make a substantive contribution to the empirical social choice literature by providing the first documented examples of monotonicity anomalies in multiwinner elections and estimating the frequency of such anomalies in real-world STV elections.

\section*{Appendix A}

In this appendix we provide a list of elections which demonstrate each type of anomaly and, for the upward, downward, and no-show anomalies, we provide a brief description of how to construct an alternative preference profile $P'$ which causes the anomaly to occur. In what follows, $P$ represents the actual preference profile and $P'$ is the modified profile. Recall that a \emph{bullet vote} for a candidate $A$ is a ballot on which $A$ is the only candidate listed on the ballot.

When listing the elections we also provide the party affiliation of each candidate. We use the following acronyms for the Scottish political parties: Conservative (Con), Green (Grn), Independent (Ind), Labour (Lab), Liberal Democrats (LD), and Scottish National Party (SNP). We note that we do not count ``Independent'' as a political party.

\subsection*{Committee Size Anomalies}

The nine elections which demonstrate this anomaly are listed below. For each election we list the year of the election first, then the council area, and finally the ward. The second winner set listed under the election name is the actual winner set which occurred in the actual election, and the first winner set demonstrates the anomaly.

\begin{itemize}
\item 2012 Dundee City, Maryfield Ward (Ward 5).

 $W(P,2)= $ $\{$Lynn (SNP), McIrvine (Lab)$\}$

 $W(P,3)= $ $\{$Cruickshank (Lab), Lynn (SNP), Melville (SNP)$\}$

\item 2012 North Lanarkshire, Cumbernauld South Ward (Ward 3). 

 $W(P,3)=$ $\{$Goldie (SNP), Graham (Lab), Homer (SNP)$\}$
 
 $W(P,4)=$ $\{$Goldie (SNP), Graham (Lab), Hogg (SNP), Muir (Lab)$\}$

\item 2017 Dumfries and Galloway, Annandale and East Eskdale Ward (Ward 12). 

$W(P,2)= $ $\{$Carruthers (Con), Male (Ind)$\}$

$W(P,3)= $ $\{$Carruthers (Con), Drynurgh (Lab), Tait (Con)$\}$

\item 2017 East Dunbartonshire, Bishopbriggs North and Campsie Ward (Ward 4). 

 $W(P,3)=$ $\{$Ferretti (SNP), Hendry (Con), Welsh (Lab)$\}$

$W(P,4)=$ $\{$Ferretti (SNP), Fischer (SNP), Hendry (Con), Pews (LD)$\}$

\item 2017 Moray, Buckie Ward (Ward 3). 

 $W(P,2)= $ $\{$Eagle (Con), McDonald (SNP)$\}$

$W(P,3)= $ $\{$Cowie (Ind), Eagle (Con), Warren (SNP)$\}$

\item 2017 West Dunbartonshire, Dumbarton Ward (Ward 3). 

 $W(P,3)=$ $\{$Black (WDuns),  Conaghan (SNP), McBride (Lab)$\}$

$W(P,4)=$ $\{$Conaghan (SNP), McBride (Lab),  McLaren (SNP), Waler (Con)$\}$

\item 2022 East Dunbartonshire, Bishopbriggs North and Campsie Ward (Ward 4). 

 $W(P,3)=$ $\{$Ferretti (SNP), McDiarmid (Lab), Pews (LD)$\}$

$W(P,4)=$ $\{$Ferretti (SNP), Hendry (Con), McDiarmid (Lab), Williamson (SNP)$\}$ 

\item 2022 City of Edinburgh, Southside/Newington Ward (Ward 15). 

 $W(P,3)=$ $\{$Burgess (Grn), Pogson (Lab), Rose (Con)$\}$

$W(P,4)=$ $\{$Burgess (Grn), Flannery (LD), Kumar (SNP), Pogson (Lab)$\}$ 

\item 2022 South Lanarkshire, East Kilbride West Ward (Ward 9). 

$W(P,2)= $ $\{$McAdams (Lab), Sloan (SNP)$\}$

$W(P,3)= $ $\{$McAdams (Lab), Salamati (SNP), Watson (Ind)$\}$

\end{itemize}

\subsection*{Upward Monotonicity Anomalies}
The 21 elections we found which demonstrate an upward monotonicity anomaly are listed below. The first line gives the year, council area, and ward of the election. The second line gives the winner set using the actual preference profile $P$ and the third line gives the new winner set when using a modified profile $P'$ after shifting the affected winning candidate up on some ballots. For each election we describe the ballots we used to create $P'$.

\begin{itemize}
\item 2012 Aberdeenshire, Stonehaven and Lower Deeside Ward (Ward 18).

$W(P,4)=$ $\{$Agnew (Con), Bellarby (LD), Christie (Lab), Clark (SNP)$\}$ 

$W(P',4)=$ $\{$Agnew (Con), Bellarby (LD),  Clark (SNP), Samways (Ind)$\}$

Create $P'$ by shifting Christie up to the first ranking on all ballots on which Shanks (Grn) is ranked first and Christie is ranked second, and five ballots on which Shanks is ranked first and Christie is ranked third.

\item 2012 Argyll Bute, Oban North and Lorn Ward (Ward 5).

$W(P,4)=$ $\{$Glen-Lee (SNP), MacDonald (Ind), MacIntyre (Ind), Robertson (Ind)$\}$

$W(P',4)=$ $\{$Glen-Lee (SNP), MacIntyre (Ind), Melville (SNP), Robertson (Ind)$\}$

Create $P'$ by changing five ballots of the form Glen-Lee $\succ$ MacDonald $\succ$ Mellville to MacDonald $\succ$ Glen-Lee $\succ$ Mellville, and two ballots of the form Robertson $\succ$ Glen-Lee $\succ$ MacDonald to Robertson $\succ$ MacDonald $\succ$ Glen-Lee.

\item 2012 Comhairle nan Eilean Siar, Sgire an Rubha Ward (Ward 5).

$W(P,3)=$ $\{$A. MacLeod (Ind), N. MacLeod (Ind), Stewart (Ind)$\}$ 

$W(P',3)=$ $\{$A. MacLeod (Ind), Nicholson (Ind),  Stewart (Ind)$\}$

Create $P'$ by shifting N. MacLeod up to the first ranking on six ballots on which MacSween (Ind) ranked first and N. MacLeod is ranked second.

\item 2012 Comhairle nan Eilean Siar, Ste\`{o}nabhagh a Tuath Ward (Ward 7).

$W(P,4)=$ $\{$MacAulay (Ind), R. MacKay (Ind), MacKenzie (Ind), Murray (SNP)$\}$ 

$W(P',4)=$ $\{$Ahmed (SNP),  R. MacKay (Ind), MacKenzie (Ind), Murray (SNP)$\}$

Create $P'$ by shifting MacAulay up to the first ranking on four ballots on which  J. MacKay (Ind) is ranked first and MacAulay is ranked second.

\item 2012 Highland, Cromarty Firth Ward (Ward 7).

$W(P,4)=$ $\{$Finlayson (Ind), Rattray (LD), Smith (SNP), Wilson (Ind)$\}$

$W(P',4)=$ $\{$Finlayson (Ind), Fletcher (SNP), Smith (SNP), Wilson (Ind)$\}$

Create $P'$ by shifting Rattray up to the first ranking on 25 ballots on which MacInness (Lab) is ranked first and Rattray is ranked second.

\item 2012 Highland, Inverness South Ward (Ward 20).

$W(P,4)=$ $\{$Caddick (LD), Crawford (Ind), Gowans (SNP), Prag (LD)$\}$

$W(P',4)=$ $\{$Caddick (LD), Crawford (Ind), Gowans (SNP), MacKenzie (Lab)$\}$

Create $P'$ by shifting Prag up to the first ranking on 8 ballots on which Boyd (SNP) is ranked first and Prag is ranked second. Furthermore, shift Prag up to first on 49 ballots on which Boyd is ranked first and Prag is ranked third.

\item 2012 Perth Kinross, Almond and Earn Ward (Ward 9)

$W(P,3)=$ $\{$Anderson (SNP), Jack (Ind), Livingstone (Con)$\}$

$W(P',3)=$ $\{$Anderson (SNP), Livingstone (Con), Lumsden (SNP)$\}$

Create $P'$ by shifting Jack up to the first ranking on one ballot on which Anderson is ranked first and Jack is ranked second.

\item 2017 Argyll and Bute, Isle of Bute Ward (Ward 8). 

$W(P,3)=$ $\{$Findlay (SNP), Moffat (Ind), Scoullar (Ind)$\}$ 

$W(P',3)=$ $\{$MacIntyre (SNP), Moffat (Ind), Wallace (Con)$\}$

Create $P'$ by taking shifting Findlay up to the first ranking on 11 ballots on which Scoullar is ranked first and Findlay is ranked second.

\item 2017 East Dunbartonshire, Lenzie \& Kirkintilloch South Ward (Ward 6). 

$W(P,3)=$ $\{$Thornton (Con), Renwick (SNP), Ackland (LD)$\}$ 

$W(P',3)=$ $\{$Thornton(Con), Renwick (SNP), Taylor (Ind)$\}$

Create $P’$ by modifying 303 ballots: 

169 ballots of the form Geekie$\succ$Ackland$\succ\dots$ modified to Ackland$\succ$Geekie$\succ\dots$ (where $\dots$ is a variety of other candidates, possibly with multiple rankings) 

51 ballots of the form Geekie$\succ$***$\succ$Ackland modified to Ackland$\succ$Geekie$\succ$*** (where *** is Scrimgeour, Sinclair, Thornton, or some combination of those three candidates)

83 ballots of the form ***$\succ$Geekie$\succ$Ackland modified to ***$\succ$Ackland$\succ$Geekie  (where *** is Scrimgeour, Sinclair, Thornton, or some combination of those three candidates)

\item 2017 City of Edinburgh, Forth Ward (Ward 4).

$W(P,4)=$ $\{$Bird (SNP), Campbell (Con), Day (Lab), Gordon (SNP)$\}$

$W(P',4)=$ $\{$Bird (SNP), Campbell (Con), Day (Lab), Mackay (Grn)$\}$

Create $P'$ by changing 43 bullet votes for Wight (LD) to ballots of the form Gordon $\succ$ Wight.

\item 2017 Fife, Kirkcaldy East Ward (Ward 12).

$W(P,3)=$ $\{$Cameron (Lab), Cavanagh (SNP), Watt (Con)$\}$ 

$W(P',3)=$ $\{$Cameron (Lab), Cavanagh (SNP), Penman (Ind)$\}$

Create $P'$ by shifting Watt up to the first ranking on six ballots on which McMahon (SNP) is ranked first.
\item 2017 Glasgow City, Govan Ward (Ward 5).

$W(P,4)=$ $\{$Bell (SNP), Dornan (SNP), Kane (Lab), Young (Grn)$\}$

$W(P',4)=$ $\{$Bell (SNP), Dornan (SNP), Kane (Lab), Shoaib (Lab)$\}$

Create $P'$ by shifting Young up to the first ranking on 72 ballots on which McCourt (Con) is ranked first, Young is ranked above Shoaib, and Young is ranked second, third, or fourth.

\item 2017 Glasgow City, Calton Ward (Ward 9).

$W(P,4)=$ $\{$Connelly (Con), Hepburn (SNP), Layden (SNP), O'Lone (Lab)$\}$

$W(P',4)=$ $\{$Hepburn (SNP), Layden (SNP), O'Lone (Lab), Rannachan (Lab)$\}$

It is difficult to describe concisely how to create $P'$; we are happy to provide the modified profile on request. In brief, shift Connelly up to the first ranking on 36 ballots on which Pike (SNP) is ranked first, and shift Connelly up to a ranking just above McLaren (Grn) on 454 ballots.

\item 2017 North Lanarkshire, Cumbernauld South Ward (Ward 3).

$W(P,4)=$ $\{$Ashraf (SNP), Goldie (SNP), Graham (Lab), Johnston (SNP)$\}$

$W(P',4)=$ $\{$Goldie (SNP), Graham (Lab), Griffin (Lab), Johnston (SNP)$\}$

Create $P'$ by shifting Ashraf up to the first ranking on five ballots on which Gibson (Con) is ranked first and Ashraf is ranked second.

\item 2017 By-Election in Perth and Kinross, Perth City South Ward (Ward 10).

$W(P,1)=$ $\{$Coates (Con)$\}$

$W(P',1)=$ $\{$Barrett (LD)$\}$

Create $P'$ by changing 151 bullet votes for Leitch (SNP) to ballots of the form Coates $\succ$ Leitch.

\item 2022 Aberdeenshire, Stonehaven and Lower Deeside Ward (Ward 18).

$W(P,4)=$ $\{$Agnew (Con), Black (SNP), Dickinson (LD), Turner (Con)$\}$

$W(P',4)=$ $\{$Agnew (Con), Black (SNP), Dickinson (LD), Simpson (Ind)$\}$

Create $P'$ by changing 15 bullet votes for Robertson (SNP) to ballots of the form Turner $\succ$ Robertson.

\item 2022 Dumfries and Galloway, Mid and Upper Nithsdale Ward (Ward 7).

$W(P,3)=$ $\{$Berretti (SNP), Dempster (Ind), Wood (Con)$\}$ 

$W(P',3)=$ $\{$Berretti (SNP), Dempster (Ind), Thornton (Con)$\}$

Create $P'$ by changing 20 bullet votes for Jamieson (Lab)  to ballots of the form Wood $\succ$ Jamieson.

\item 2022 City of Edinburgh, Inverleith Ward (Ward 5).

$W(P,4)=$ $\{$Bandel (Grn), Mitchell (Con), Nicolson (SNP), Osler (LD)$\}$

$W(P',4)=$ $\{$Mitchell (Con), Munro-Brian (Lab), Nicolson (SNP), Osler (LD)$\}$

Create $P'$ by shifting Bandel up to the first ranking on 12 ballots on which Wood (LD) is ranked first and Bandel is ranked second.

\item 2022 Fife, Kircaldy North Ward (Ward 10).

$W(P,3)=$ $\{$Leslie (Con), Lindsay (SNP), Ross (Lab)$\}$ 

$W(P',3)=$ $\{$Lindsay (SNP), Ross (Lab), Smart (Lab)$\}$

Create $P'$ by changing 93 ballots of the form Walsh (SNP) $\succ$ Lindsay to ballots of the form Leslie $\succ$ Walsh  $\succ$ Lindsay.

\item 2022 Glasgow City, Garscadden/Scotstounhill Ward (Ward 13).

$W(P,4)=$ $\{$Butler (Lab), Cunningham (SNP), Mitchell (SNP), Murray (Lab)$\}$ 

$W(P',4)=$ $\{$Butler (Lab), Cunningham (SNP), Murray (Lab), Ugbah (SNP)$\}$

Create $P'$ by changing 18 ballots of the form Hamelink (Grn) $\succ$ Cunningham $\succ$ Mitchell $\succ$ Ugbah to ballots of the form
Mitchell $\succ$ Hamelink (Grn) $\succ$ Cunningham $\succ$ Ugbah.

\item 2022 Highland, Inverness West Ward (Ward 13).

$W(P,3)=$ $\{$Boyd (SNP), Graham (LD), MacKintosh (Grn)$\}$ 

$W(P',3)=$ $\{$Boyd (SNP), Fraser (Lab), Graham (LD)$\}$

Create $P'$ by shifting MacKintosh up to the first ranking on seven ballots on which Forbes (Con) is ranked first and MacKintosh is ranked second.

\item 2022 Orkney Islands, East Mainland South Ronaldsay and Burray Ward (Ward 5).

$W(P,3)=$ $\{$Moar (Ind), Peace (Ind), Skuse (Ind)$\}$ 

$W(P',3)=$ $\{$Peace (Ind), Rickards (Ind), Skuse (Ind)$\}$

Create $P'$ by changing three bullet votes for Page (Grn)  to Moar $\succ$ Page. Furthermore, shift Moar up to the first ranking on 36 ballots on which Page is ranked first and Moar is listed on the ballot above Rickards.

\item 2022 South Lanarkshire, Rutherglen Central and North Ward (Ward 12).

$W(P,3)=$ $\{$Calikes (SNP), Cowan (SNP), Lennon (Lab)$\}$ 

$W(P',3)=$ $\{$Calikes (SNP), Lennon (Lab), McGinty (Lab)$\}$

Create $P'$ by changing all 26 ballots of the form Fox (Con) $\succ$ McGinty $\succ$ Lennon to ballots of the form Cowan $\succ$ Fox $\succ$ McGinty $\succ$ Lennon.

\end{itemize}

\subsection*{Downward Monotonicity Anomalies}
The 15 elections we found which demonstrate a downward monotonicity anomaly are listed below. The first line gives the year, council area, and ward of the election. The second line gives the winner set using the actual preference profile $P$ and the third line gives the new winner set when using a modified profile $P'$ after shifting the affected losing candidate down on some ballots. For each election we describe the ballots we used to create $P'$.

\begin{itemize}

\item 2012 Comhairle nan Eilean Siar, Ste\`{o}nabhagh a Tuath Ward (Ward 7).

$W(P,4)=$ $\{$MacAulay (Ind), R. MacKay (Ind), MacKenzie (Ind), Murray (SNP)$\}$ 

$W(P',4)=$ $\{$Ahmed (SNP),  R. MacKay (Ind), MacKenzie (Ind), Murray (SNP)$\}$

Create $P'$ by shifting Ahmed down one ranking on 12 ballots on which Ahmed is ranked first and Campbell (Ind) is ranked second.

\item 2012 Highland, Cromarty Firth Ward (Ward 7).

$W(P,4)=$ $\{$Finlayson (Ind), Rattray (LD), Smith (SNP), Wilson (Ind)$\}$ 

$W(P',4)=$ $\{$Fletcher (SNP), Finlayson (Ind), Smith (SNP), Wilson (Ind)$\}$

Create $P'$ by shifting Fletcher down on ranking on 9 ballots on which Fletcher is ranked first and McCaffery (Ind) is ranked second.

\item 2012 Highland, Inverness South Ward (Ward 20).

$W(P,4)=$ $\{$Caddick (LD), Crawford (Ind), Gowans (SNP), Prag (LD)$\}$

$W(P',4)=$ $\{$Boyd (SNP), Caddick (LD), Crawford (Ind), Gowans (SNP)$\}$

Create $P'$ by shifting Boyd down one ranking on the 4 ballots on which Boyd is ranked first and Bonsor (Con) is ranked second.


\item 2017 Aberdeenshire, Ellon and District Ward (Ward 9).

$W(P,4)=$ $\{$Davidson (LD), Kahanov-Kloppert (SNP), Owen (Con), Thomson (SNP)$\}$

$W(P',4)=$ $\{$Davidson (LD), Morgan (Lab), Owen (Con), Thomson (SNP)$\}$

Create $P'$ by shifting Morgan down one ranking on 17 ballots on which Morgan is ranked first and Davidson is ranked second.

\item 2017 Argyll and Bute, Isle of Bute Ward (Ward 8). 

$W(P,3)=$ $\{$Findlay (SNP), Moffat (Ind), Scoullar (Ind)$\}$ 

$W(P',3)=$ $\{$MacIntyre (SNP), Moffat (Ind), Wallace (Con)$\}$

Create $P'$ by shifting Wallace down one ranking on 12 ballots on which Wallace is ranked first and Gillies (Ind) is ranked second.

\item 2017 North Ayrshire, Saltcoats Ward (Ward 9).

$W(P,3)=$ $\{$McClung (SNP), McNicol (Ind), Montgomerie (Lab)$\}$ 

$W(P',3)=$ $\{$Clydesdale (Con), McClung (SNP), Montgomerie (Lab)$\}$

Create $P'$ by changing 41 bullet votes for Clydesdale to Bianchini (SNP) $\succ$ Clydesdale.

\item 2017 North Lanarkshire, Mossend and Holytown Ward (Ward 16).

$W(P,3)=$ $\{$Baird (SNP), McNally (Lab), Reddin (Lab)$\}$ 

$W(P',3)=$ $\{$Baird (SNP), Cunningham (Con), McNally (Lab)$\}$

Create $P'$ by shifting Cunningham down one ranking on 11 ballots on which Cunningham is ranked first and Clarkson (SNP) is ranked second. Furthermore, change the 9 ballots of the form Cunningham $\succ$ Baird $\succ$ Clarkson to Baird $\succ$ Clarkson $\succ$ Cunningham.

\item 2017 North Lanarkshire, Murdostoun Ward (Ward 20).

$W(P,4)=$ $\{$McKendrick (Ind), McManus (SNP), Roarty (Lab), Shevlin (Lab)$\}$

$W(P',4)=$ $\{$MacKenzie (Con), McKendrick (Ind), McManus (SNP), Roarty (Lab)$\}$

Create $P'$ by changing all ballots with MacKenzie ranked first and Millar (UKIP) ranked second so that Millar is ranked first and MacKenzie second. Furthermore, change 38 bullet votes for MacKenzie to Millar $\succ$ MacKenzie.

\item 2017 Renfrewshire, Paisley Southeast Ward (Ward 6).

$W(P,3)=$ $\{$Devine (Lab), Mack (Ind), McGurk (SNP)$\}$ 

$W(P',3)=$ $\{$Devine (Lab), Fulton (Con), McGurk (SNP)$\}$

Create $P'$ by changing 18 bullet votes for Fulton to Swanson (SNP) $\succ$ Fulton. Alternatively, change these ballots to bullet votes for Swanson.

\item 2017 Stirling, Dunblane and Bridge of Allan Ward (Ward 3).

$W(P,4)=$ $\{$Dodds (Con), Houston (SNP), Majury (Con), Tollemache (Grn)$\}$

$W(P',4)=$ $\{$Dodds (Con), Houston (SNP), Majury (Con), Robbins (Lab)$\}$

Create $P'$ by changing the 144 bullet votes for Robbins to Hunter (SNP) $\succ$ Robbins. Alternatively, change the 144 bullet votes for Robbins to bullet votes for Hunter.

\item 2022 Aberdeen City, George Street/Harbour Ward (Ward 8).

$W(P,4)=$ $\{$Bouse (LD), Henrickson (SNP), Hutchison (SNP), Macdonald (Lab)$\}$

$W(P',4)=$ $\{$Henrickson (SNP), Hutchison (SNP), Ingerson (Grn), Macdonald (Lab)$\}$

Create $P'$ by changing 32 ballots for Ingerson  to Painter (Con) $\succ$ Ingerson.

\item 2022 Aberdeenshire, Mid-Formartine Ward (Ward 8).

$W(P,4)=$ $\{$Hassan (LD), Johnston (Ind), Nicol (SNP), Ritchie (Con)$\}$

$W(P',4)=$ $\{$Hassan (LD), Johnston (Ind), Powell (Con), Ritchie (Con)$\}$

Create $P'$ by changing 25 bullet votes for Powell to Hutchison (SNP) $\succ$ Powell.

\item 2022 Argyll and Bute, Isle of Bute Ward (Ward 8). 

$W(P,3)=$ $\{$Kennedy-Boyle (SNP), McCabe (Ind), Wallace (Con)$\}$ 

$W(P',3)=$ $\{$Kennedy-Boyle (SNP), McCabe (Ind), Moffat (Ind)$\}$

Create $P'$ by changing 41 bullet votes for Moffat to Stuart (Grn) $\succ$ Moffat. 

\item 2022 City of Edinburgh, Inverleith Ward (Ward 5).

$W(P,4)=$ $\{$Bandel (Grn), Mitchell (Con), Nicolson (SNP), Osler (LD)$\}$

$W(P',4)=$ $\{$Mitchell (Con),  Nicolson (SNP), Osler (LD), Wood (LD))$\}$

Create $P'$ changing one ballot of the form Wood $\succ$ Nicolson to Nicolson $\succ$ Wood.

\item 2022 Falkirk, Grangemouth Ward (Ward 2).

$W(P,3)=$ $\{$Balfour (SNP), Nimmo (Lab), Spears (Ind)$\}$ 

$W(P',3)=$ $\{$Balfour (SNP), Haston (SNP), Nimmo (Lab)$\}$

Create $P'$ by shifting Haston down one ranking on four ballots on which Haston is ranked first and Bryson (Con) is ranked second. Furthermore, change all 17 ballots of the form Balfour $\succ$ Haston $\succ$ Bryson  to Balfour $\succ$ Bryson $\succ$ Haston. 

\item 2022 Glasgow City, Patrick East/Kelvindale Ward (Ward 23).

$W(P,4)=$ $\{$Anderson (Grn), Brown (Lab), Johnstone (Lab), McLean (SNP)$\}$

$W(P',4)=$ $\{$Anderson (Grn), Asghar (Con), Brown (Lab), McLean (SNP)$\}$

Create $P'$ by changing 147 bullet votes for Asghar and change them to bullet votes for Wilson (SNP).






\item 2022 Perth and Kinross, Highland Ward (Ward 4)

$W(P,3)=$ $\{$Duff (Con), McDade (Ind), Williamson (SNP)$\}$ 

$W(P',3)=$ $\{$Duff (Con), McDade (Ind), Murray (SNP)$\}$

Create $P'$ by shifting Murray down one ranking on 37 ballots with Murray ranked first and McDougall (Grn) ranked second.

\end{itemize}

\subsection*{No-Show Anomalies}

The 15 elections we found which demonstrate a no-show anomaly are listed below. The first line gives the year, council area, and ward of the election. The second line gives the winner set using the actual preference profile $P$ and the third line gives the new winner set when using a modified profile $P'$ after shifting the affected losing candidate down on some ballots. For each election we describe the ballots we used to create $P'$.

\begin{itemize}

\item 2012 Aberdeenshire, Stonehaven and Lower Deeside Ward (Ward 18).

$W(P,4)=$ $\{$Agnew (Con), Bellarby (LD), Christie (Lab), Clark (SNP)$\}$ 

$W(P',4)=$ $\{$Agnew (Con), Bellarby (LD), Clark (SNP), Samways (Ind)$\}$

Create $P'$ by removing 16 ballots on which Michie (Ind) is ranked first, Samways is ranked third or fourth, and Christie does not appear on the ballot.

\item 2012 Argyll Bute, Oban North and Lorn Ward (Ward 5).

$W(P,4)=$ $\{$Glen-Lee (SNP), MacDonald (Ind), MacIntyre (Ind), Robertson (Ind)$\}$

$W(P',4)=$ $\{$Glen-Lee (SNP), MacIntyre (Ind), Melville (SNP), Robertson (Ind)$\}$

Create $P'$ by removing 8 ballots of the form Glen-Lee $\succ$ Melville.

\item 2012 Comhairle nan Eilean Siar, Sgire an Rubha Ward (Ward 5).

$W(P,3)=$ $\{$A. MacLeod (Ind), N. MacLeod (Ind), Stewart (Ind)$\}$ 

$W(P',3)=$ $\{$A. MacLeod (Ind), Nicholson (Ind),  Stewart (Ind)$\}$

Create $P'$ by removing 4 ballots of the form MacSween $\succ$ Nicholson. Alternatively, we can remove 4 ballots of the form  MacSween $\succ$ Nicholson $\succ$ Stewart; it seems unambiguous that these 4 voters do not want N. MacLeod in the winner set.

\item 2012 Comhairle nan Eilean Siar, Ste\`{o}nabhagh a Tuath Ward (Ward 7).

$W(P,4)=$ $\{$MacAulay (Ind), R. MacKay (Ind), MacKenzie (Ind), Murray (SNP)$\}$ 

$W(P',4)=$ $\{$Ahmed (SNP),  R. MacKay (Ind), MacKenzie (Ind), Murray (SNP)$\}$

Create $P'$ by removing removing the following four ballots.

J. MacKay $\succ$ R. MacKay $\succ$ G. Murray $\succ$ Ahmed

J. MacKay $\succ$ R. MacKay $\succ$ G. Murray $\succ$ Ahmed

J. MacKay $\succ$ Ahmed $\succ$ G. Murray $\succ$ R. MacKay

J. MacKay $\succ$ G. Murray $\succ$ Ahmed $\succ$ R. MacKay $\succ$ MacAulay $\succ$ Campbell

\item 2012 East Ayrshire, Kilmarnock South Ward (Ward 5)

$W(P,3)=$ $\{$Knapp (Lab), Ross (SNP), Todd (SNP)$\}$

$W(P,3)=$ $\{$Knapp (Lab), Scott (Lab), Todd (SNP)$\}$

Create $P'$ by removing all ballots on which Todd is ranked first, Scott is ranked second, and Ross is not ranked in the top three.

\item 2012 Falkirk, Bo'ness and Blackness Ward (Ward 1)

$W(P,3)=$ $\{$Mahoney (Lab), Ritchie (SNP), Turner (SNP)$\}$

$W(P',3)=$ $\{$Aitchison (Lab), Mahoney (Lab), Ritchie (SNP)$\}$

Create $P'$ by removing all ballots of the following forms: Ritchie $\succ$ Aitchison; Ritchie $\succ$ Aitchison $\succ$ Mahoney; Ritchie $\succ$ Mahoney $\succ$ Aitchison.

\item 2012 Highland, Thurso Ward (Ward 2)

$W(P,3)=$ $\{$MacKay (Ind), Rosie (Ind), Saxon (Lab)$\}$

$W(P',3)=$ $\{$MacKay (Ind), Saxon (Lab), Smith (SNP)$\}$

Create $P'$ by removing all ballots of the form MacKay $\succ$ Saxon $\succ$ Smith and  MacKay $\succ$ Smith $\succ$ Saxon. Furthermore, remove 15 ballots of the form  MacKay $\succ$ Smith.

\item 2012 Highland, Cromarty Firth Ward (Ward 7)

$W(P,4)=$ $\{$Finlayson (Ind), Rattray (LD), Smith (SNP), Wilson (Ind)$\}$ 

$W(P',4)=$ $\{$Finlayson (Ind), Fletcher (SNP), Smith (SNP), Wilson (Ind)$\}$ 

Create $P'$ by removing 17 ballots of the form Smith $\succ$ Fletcher.

\item 2012 Highland, Dingwall and Seaforth Ward (Ward 9)

$W(P,4)=$ $\{$MacKenzie (SNP), McKinnon (Ind), MacLean (LD), Paterson (Ind)$\}$ 

$W(P',4)=$ $\{$Erskine (Lab), MacKenzie (SNP), MacLean (LD), Paterson (Ind)$\}$ 

Create $P'$ by removing all 58 ballots with Paterson ranked first, Erskine ranked second, and McKinnon not ranked in the top 4. In addition, remove four ballots of the form MacLean $\succ$ MacKenzie $\succ$ Erskine.

$W(P,3)=$ $\{$Gowans (SNP), Jarvis (Lab), Shand (SNP)$\}$

\item 2012 Highland, Inverness South Ward (Ward 20)

$W(P,4)=$ $\{$Caddick (LD), Crawford (Ind), Gowans (SNP), Prag (LD)$\}$ 

$W(P,4)=$ $\{$Boyd (SNP), Caddick (LD), Crawford (Ind), Gowans (SNP)$\}$ 

Create $P'$ by removing 21 ballots with Caddick ranked first, Boyd ranked in the top 4, and Prag not ranked in the top 4.

\item 2012 Moray, Heldon and Laich Ward (Ward 5)

$W(P,4)=$ $\{$McGillivray (Ind), Ralph (SNP), Tuke (Ind), Wright (Con)$\}$ 

$W(P',4)=$ $\{$McGillivray (Ind), Ralph (SNP), Stewart (SNP), Wright (Con)$\}$ 

Create $P'$ by removing 57 ballots with McGillivray ranked first, Stewart ranked second, and Tuke not ranked in the top four.

\item 2012 Moray, Elgin City North Ward (Ward 6)

$W(P,3)=$ $\{$Gowans (SNP), Jarvis (Lab), Shand (SNP)$\}$

$W(P',3)=$ $\{$Brown (Con), Jarvis (Lab), Shand (SNP)$\}$

Create $P'$ by removing the two ballots of the form Shand $\succ$ Brown $\succ$ Margach $\succ$ Gowans $\succ$ Jarvis.

\item 2012 Perth Kinross, Almond and Earn Ward (Ward 9)

$W(P,3)=$ $\{$Anderson (SNP), Jack (Ind), Livingstone (Con)$\}$

$W(P',3)=$ $\{$Anderson (SNP), Livingstone (Con), Lumsden (SNP)$\}$

Create $P'$ by removing one ballot of the form Anderson $\succ$ Lumsden $\succ$ Dundas $\succ$ Hayton $\succ$ Livingstone $\succ$ Jack. (This election contains six candidates, so this voter is ranking Jack last.)

\item 2017 City of Edinburgh, Forth Ward (Ward 4).

$W(P,4)=$ $\{$Bird (SNP), Campbell (Con), Day (Lab), Gordon (SNP)$\}$

$W(P',4)=$ $\{$Bird (SNP), Campbell (Con), Day (Lab), Mackay (Grn)$\}$

Create $P'$ by removing 46 ballots of the form Wight (LD) $\succ$ Mackay.

\item 2017 Aberdeenshire, Ellon and District (Ward 9)

$W(P,4)=$ $\{$Davidson (LD), Kahanov-Kloppert (SNP), Owen (Con), Thomson (SNP)$\}$

$W(P',4)=$ $\{$Davidson (LD), Morgan (Lab), Owen (Con), Thomson (SNP)$\}$

Create $P'$ by removing 85 ballots of the form Owen $\succ$ Morgan $\succ$ Davidson.

\item 2017 Dumfries and Galloway, Ninth Ward (Ward 9)

$W(P,4)=$ $\{$Campbell (SNP), Johnstone (Con), Martin (Lab), Murray (Lab)$\}$

$W(P',4)=$ $\{$Campbell (SNP), Johnstone (Con),  Murray (Lab), Slater (Ind)$\}$

Create $P'$ by removing 23 ballots with Murray ranked first, Slater ranked second, and Martin not ranked in the top four.

\item 2017 East Ayrshire, Irvine Valley Ward (Ward 6)

$W(P,3)=$ $\{$Cogley (Rub), Mair (Lab), Whitham (SNP)$\}$

$W(P',3)=$ $\{$Cogley (Rub), McFadzean (Con), Whitham (SNP)$\}$

Create $P'$ by removing 26 ballots with Whitham ranked first, Young (SNP) ranked second, and McFadzean ranked third.

\item 2017 Edinburgh, Colinton/Fairmilehead (Ward 8)

$W(P,3)=$ $\{$Arthur (Lab), Doggart (Con), Rust (Con)$\}$

$W(P',3)=$ $\{$Arthur (Lab), Lewis (SNP), Rust (Con)$\}$

Create $P'$ by removing all ballots on which Rust is ranked first, Lewis ranked second, and Doggart is not ranked in the top three.

\item 2017 Fife, Kirkcaldy East Ward (Ward 12).

$W(P,3)=$ $\{$I. Cameron (Lab), Cavanagh (SNP), Watt (Con)$\}$ 

$W(P',3)=$ $\{$I. Cameron (Lab), Cavanagh (SNP), Penman (Ind)$\}$

Create $P'$ by removing 7 ballots of the form McMahon (SNP) $\succ$ Cavanagh $\succ$ Penman.

\item 2017 Glasgow City, Calton Ward (Ward 9). 

$W(P,4) =$ $\{$Hepburn (SNP), Layden (SNP), O’Lone (Lab), Connelly (Con)$\}$

$W(P’,4) =$ $\{$Hepburn (SNP), Layden (SNP), O’Lone (Lab), Rannachan (Lab)$\}$

Create $P’$ by removing 49 ballots:
 
34 ballots of the form Pike $\succ$ *** $\succ$ Rannachan $\succ\dots$  (where *** is either no one or a variety of candidates that are not Connelly, and $\dots$ is either no one or a variety of candidates possibly including Connelly) 

8 ballots of the form Hepburn $\succ$ *** $\succ$ Pike $\succ$ *** $\succ$ Rannachan$\succ\dots$ (where *** is either no one or a variety of candidates that are not Connelly, and $\dots$ is either no one or a variety of candidates possibly including Connelly) 

7 ballots of the form McLaren $\succ$ Pike $\succ$ *** $\succ$ Rannachan $\succ\dots$ (where *** is either no one or a variety of candidates that are not Connelly, and $\dots$ is either no one or a variety of candidates possibly including Connelly) 

\item 2017 Moray, Heldon and Laich Ward  (Ward 5)

$W(P,4) =$ $\{$Allan (Con), Cowe (Ind), Edwards (Ind), Patience (SNP)$\}$

$W(P',4) =$ $\{$Allan (Con), Cowe (Ind),  Patience (SNP), Slater (Ind)$\}$

Create $P'$ by removing all ballots of the form Cowe $\succ$ Slater and all ballots of the form Cowe $\succ$ Slater $\succ$ Allan.

\item 2017 North Lanarkshire, Cumbernauld South Ward (Ward 3).

$W(P,4)=$ $\{$Ashraf (SNP), Goldie (SNP), Graham (Lab), Johnston (SNP)$\}$

$W(P',4)=$ $\{$Goldie (SNP), Graham (Lab), Griffin (Lab), Johnston (SNP)$\}$

Create $P'$ be removing six ballots of the form Gibson $\succ$ Griffin $\succ$ Graham $\succ$ Homer.

\item 2017 North Lanarkshire, Coatbridge South Ward (Ward 11)

$W(P,4)=$ $\{$Carragher (SNP), Castles (Lab), Encinias (Lab), MacGregor (SNP)$\}$

$W(P,4)=$ $\{$Brooks (IANL), Carragher (SNP), Castles (Lab), MacGregor (SNP)$\}$

Create $P'$ by removing the single ballot of the form Castles $\succ$ Brooks $\succ$ Higgins $\succ$ Cameron $\succ$ Somers $\succ$ Encinias.

\item 2017 By-Election in Perth and Kinross, Perth City South Ward (Ward 10).

$W(P,1)=$ $\{$Coates (Con)$\}$

$W(P',1)=$ $\{$Barrett (LD)$\}$

Create $P'$ by removing 82 ballots of the form Leitch $\succ$ Barrett, 53 ballots of the form  Leitch $\succ$ Barrett $\succ$ MacLachlan, 5 ballots of the form  Leitch $\succ$ Barrett $\succ$ MacLachlan $\succ$ Baykal, and 11 ballots of the form Leitch $\succ$ Barrett $\succ$ MacLachlan $\succ$ Baykal $\succ$ Coates.

\item 2017 Shetland, Lerwick South Ward (Ward 7).

$W(P,4)=$ $\{$Campbell (Ind), Smith (Ind), Westlake (Ind), Wishart (Ind)$\}$

$W(P',4)=$ $\{$Campbell (Ind), Smith (Ind), Valente (Ind), Wishart (Ind)$\}$

Create $P'$ by removing one ballot of the form Campbell $\succ$ Valente  $\succ$ Smith  $\succ$ Wishart  $\succ$ Westlake.

\item 2017 West Dunbartonshire, Leven Ward (Ward 2)

$W(P,4)=$ $\{$Bollan (WDuns), Dickson (SNP), McAllister (SNP), Millar (Lab)$\}$

$W(P',4)=$ $\{$Bollan (WDuns), Dickson (SNP), McGinty (Lab), Millar (Lab)$\}$

Create $P'$ by removing 20 ballots in which Dickson is ranked first, McGinty is ranked second, and McAllistar is not ranked in the top four.

\item 2022 Aberdeenshire, Stonehaven and Lower Deeside Ward (Ward 18).

$W(P,4)=$ $\{$Agnew (Con), Black (SNP), Dickinson (LD), Turner (Con)$\}$ 

$W(P',4)=$ $\{$Agnew (Con), Black (SNP), Dickinson (LD), Simpson (Ind)$\}$

Create $P'$ by removing 15 ballots in which Robertson is ranked first, Simpson is ranked second or third, and Turner does not appear on the ballot or is ranked 8th (out of 8 candidates). Also, remove one ballot on which Black is ranked first, Simpson is ranked third, and Turner does not appear on the ballot.

\item 2022 Argyll Bute, Oban South and the Isles Ward (Ward 4)

$W(P,4)=$ $\{$Hampsey (Con), Hume (SNP), Kain (Ind), Lynch (SNP)$\}$ 

$W(P,4)=$ $\{$Hampsey (Con), Hume (SNP),  Lynch (SNP), Meyer (Grn)$\}$ 

Create $P'$ by removing the four ballots of the form Hampsey $\succ$ Boswell (LD) $\succ$ Meyer.

\item 2022 Argyll Bute, Isle of Bute (Ward 8)

$W(P,3)=$ $\{$Kennedy-Boyle (SNP), McCabe (Ind), Wallace (Con)$\}$ 

$W(P',3)=$ $\{$Kennedy-Boyle (SNP), McCabe (Ind), Moffat (Ind)$\}$ 

Create $P'$ by removing 10 ballots of the form McCabe $\succ$ Moffat $\succ$ Kennedy-Boyle.

\item 2022 Dumfries and Galloway, Mid and Upper Nithsdale Ward (Ward 7).

$W(P,3)=$ $\{$Berretti (SNP), Dempster (Ind), Wood (Con)$\}$ 

$W(P',3)=$ $\{$Berretti (SNP), Dempster (Ind), Thornton (Con)$\}$

Create $P'$ by removing 23 ballots on which Jamieson is ranked first, Thornton is ranked second or third, and Wood either doesn't appear on the ballot or is ranked 5th (out of 5 candidates).

\item 2022 City of Edinburgh, Pentland Hills Ward (Ward 2)

$W(P,4)=$ $\{$Bruce (Con), Gardiner (SNP), Glasgow (SNP), Jenkinson (Lab)$\}$

$W(P',4)=$ $\{$Bruce (Con), Gardiner (SNP), Gilchrist (Con), Jenkinson (Lab)$\}$

Create $P'$ by removing 20 ballots on which Gardiner is ranked first, Gilchrist is ranked in the top three, and Glasgow is not ranked in the top four.

\item 2022 City of Edinburgh, Inverleith Ward (Ward 5).

$W(P,4)=$ $\{$Bandel (Grn), Mitchell (Con), Nicolson (SNP), Osler (LD)$\}$

$W(P',4)=$ $\{$Mitchell (Con), Munro-Brian (Lab), Nicolson (SNP), Osler (LD)$\}$

Create $P'$ by taking removing 14 ballots on which Wood is ranked first, Munro-Brian is ranked second or third, and Bandel is not listed on the ballot.

\item 2022 Fife, Cowdenbeath Ward (Ward 7).

$W(P,4)=$ $\{$Bain (SNP), Campbell (Lab), Robb (SNP), Watt (Con)$\}$

$W(P',4)=$ $\{$Bain (SNP), Campbell (Lab), Guichan (Lab), Watt (Con)$\}$

Create $P'$ by removing 17 ballots of the form Bain $\succ$ Campbell $\succ$ Guichan.

\item 2022 Fife, Kircaldy North Ward (Ward 10).

$W(P,3)=$ $\{$Leslie (Con), Lindsay (SNP), Ross (Lab)$\}$ 

$W(P',3)=$ $\{$Lindsay (SNP), Ross (Lab), Smart (Lab)$\}$

Create $P'$ by removing 93 ballots on which Walsh is ranked first and Smart is ranked above Leslie. In this case, we cannot find a subset of ballots to remove in which Smart is always ranked in the top three.

\item 2022 Glasgow City, Garscadden/Scotstounhill Ward (Ward 13).

$W(P,4)=$ $\{$Butler (Lab), Cunningham (SNP), Mitchell (SNP), Murray (Lab)$\}$ 

$W(P',4)=$ $\{$Butler (Lab), Cunningham (SNP), Murray (Lab), Ugbah (SNP)$\}$

Create $P'$ by removing 19 ballots in which Hamelink is ranked first, Ugbah is ranked second or third, and Mitchell does not appear on the ballot.

\item 2022 Highland, Inverness West Ward (Ward 13).

$W(P,3)=$ $\{$Boyd (SNP), Graham (LD), MacKintosh (Grn)$\}$ 

$W(P',3)=$ $\{$Boyd (SNP), Fraser (Lab), Graham (LD)$\}$

Create $P'$ by taking removing the following two ballots.

Forbes $\succ$ Fraser $\succ$ Boyd

Forbes $\succ$ Boyd $\succ$ Fraser $\succ$ Graham $\succ$ Forsyth $\succ$ MacKintosh

We can also demonstrate a no-show anomaly in this election by removing three ballots of the form Forbes $\succ$ Fraser $\succ$ McDonald, which also changes the winner set to $\{$Boyd (SNP), Fraser (Lab), Graham (LD)$\}$.

\item 2022 North Lanarkshire, Motherwell West (Ward 17)

$W(P,3)=$ $\{$Crichton (SNP), Kelly (Lab), Nolan (Con)$\}$ 

$W(P,3)=$ $\{$Crichton (SNP), Evans (SNP), Kelly (Lab)$\}$ 

Remove the ballots of the following forms. The number in parentheses denotes the number of the type of ballot.

(3) Kelly $\succ$ Evans

(6) Kelly $\succ$ Evans $\succ$ Crichton

(4) Kelly $\succ$ Evans $\succ$ McCann

(3) Kelly $\succ$ Evans $\succ$ Miller

(1) Kelly $\succ$ Evans $\succ$ Crichton $\succ$ Miller $\succ$ McCann

(2) Kelly $\succ$ Evans $\succ$ McCann $\succ$ Crichton $\succ$ Miller $\succ$ Nolan

(1) Kelly $\succ$ Evans $\succ$ Miller $\succ$ Crichton

\item 2022 South Lanarkshire, Rutherglen Central and North Ward (Ward 12).

$W(P,3)=$ $\{$Calikes (SNP), Cowan (SNP), Lennon (Lab)$\}$ 

$W(P',3)=$ $\{$Calikes (SNP), Lennon (Lab), McGinty (Lab)$\}$

Create $P'$ by removing 16 ballots of the form Fox $\succ$ McGinty $\succ$ Adebo.

\item 2022 West Lothian, Whitburn and Blackburn (Ward 7)

$W(P,4)=$ $\{$J. Dickson (SNP), M. Dickson (SNP), Paul (Lab), Sullivan (Lab)$\}$ 

$W(P',4)=$ $\{$J. Dickson (SNP), M. Dickson (SNP), Fairbairn (Con), Paul (Lab)$\}$ 

Create $P'$ by removing 10 ballots of the form Paul $\succ$ Fairbairn.

\end{itemize}

\end{document}